\documentclass[prd, 10pt,aps,nofootinbib, floatfix, notitlepage,twocolumn]{revtex4-1}
\usepackage{amsmath,amsfonts,amssymb}
\usepackage{mathrsfs}
\usepackage{graphicx}
\usepackage[english]{babel} 
\usepackage{url} 
\usepackage{color}
\usepackage[utf8]{inputenc} 
\usepackage{float}
\usepackage[colorlinks=true,citecolor=blue,urlcolor=blue]{hyperref}
\usepackage{booktabs}
\usepackage{subfigure}
\usepackage{enumerate}

\usepackage[T1]{fontenc}

\newcommand{\minew}[1]{{\color{black}{#1}}}
\newcommand{\miold}[1]{\iffalse{#1}\fi}

\begin{document}


\title{Limit on the dark matter mass from its interaction with photons}


\author{Zhihuan Zhou}
\email{11702005@mail.dlut.edu.cn}
\altaffiliation[Also at ]{}
\author{Gang Liu}%
\author{Yuhao Mu}%
\author{Lixin Xu}%
\email{lxxu@dlut.edu.cn}
\affiliation{Institute of Theoretical Physics \\
	School of Physics \\
	Dalian University of Technology \\
	Dalian 116024, People's Republic of China}%




\date{\today}

\begin{abstract}
In this work, we explore the phenomenology of generalized dark matter (GDM) which interacts with photons ($\gamma$).
We assume that DM establishes elastic scattering with $\gamma$ when it has already become nonrelativistic,
otherwise the abundance of DM today is disfavored by current observations.
Within this scenario, the equation of state (EoS) of DM is determined by its mass ($m_\chi$) and the DM-$\gamma$ scattering cross-section.
The distinctive imprints of a nonzero EoS of DM on CMB angular power spectrum allow us to set a lower limit on $m_\chi$ with \emph{Planck} 2018 data alone, i.e., $m_{\chi} > 8.7$ keV at $95\%$ C.L. 
In the study of cosmic concordance problems, we find that the GDM scenario preserves the sound horizon ($r_s(z_*)$) predicted in the fiducial $\Lambda$CDM model, and thus does not solve the $H_0$ tension.
When performing the joint analysis of \emph{Planck}+LSS datasets, the best-fit $S_8= 0.785\pm 0.017$ closely matches the given $S_8$ prior. This suggests that the GDM scenario can be counted as a viable candidate to restore the $S_8$ ($\sigma_{8}$) tension.

\end{abstract}


\maketitle

\section{Introduction}
Despite that tremendous progresses in observational cosmology \cite{SupernovaSearchTeam:1998fmf,Aghanim:2018eyx,Planck:2019nip,Abbott:2017wau,Alam:2016hwk} have allowed us to determine the 
standard $\Lambda$CDM model
with great precision, the origin and composition of dark matter (DM) and dark energy (DE) \cite{Li_2011} which comprise most of the energy content in the universe remain unknown. 
In the standard picture, the cold dark matter (CDM) consists of non-interacting or weakly-interacting massive
particles (WIMPS) of mass $10\:\rm{GeV} \lesssim m_\chi \lesssim 1 \rm{TeV}$, which currently could only be observed through its gravitational effect \cite{Persic:1995ru,Clowe:2006eq,Harvey:2015hha}. 
Although the CDM paradigm is in good agreement with a large bunch of experiments, we still
lack of direct evidence for its existence and understanding of its particle nature.
Hence, extensive work has been dedicated to determine the mass of DM \cite{griest1990unitarity, Boehm:2000gq, Boyarsky:2008ju,Kavanagh:2013wba, Berlin:2017ftj,DarkSide:2018bpj}. 

With the increase in precision of cosmological measurement, the standard picture is faced with the cosmic concordance problems. 
The most famous discrepancy are known as the $H_0$ tension \cite{Perivolaropoulos:2021jda,Verde:2019ivm,Aghanim:2018eyx,Verde:2019ivm,Perivolaropoulos:2021jda,DiValentino:2021izs} and the $\sigma_{8}$ ($S_8$)
tension \cite{Abbott:2017wau,Hildebrandt:2016iqg,Hildebrandt:2018yau,HSC:2018mrq}. Moreover, observations on galaxy scales indicates that $\Lambda$CDM model are faced with serval problems\cite{Bellazzini:2013foa,TULIN20181}
in describing structure at small scales Including the core-cusp problem, the diversity problem for rotation curves, and the missing satellites problem
(see Ref. \cite{Salucci:2018hqu} for a review).
These issues have motivated people to explore the physics beyond the standard CDM paradigm. 
One important aspect is the possible interaction of DM with standard model particles like baryons \cite{Barkana:2018lgd,Munoz:2017qpy, Ali-Haimoud:2015pwa, Munoz:2015bca, Gluscevic:2017ywp, Slatyer:2018aqg, Xu:2018efh}, neutrinos \cite{Bringmann:2013vra,Audren:2014lsa,Horiuchi:2015qri,DiValentino:2017oaw,Pandey:2018wvh,Choi:2019ixb,Stadler:2019dii}, or photons \cite{Wilkinson:2013kia,Stadler:2018jin, Becker:2020hzj, Weiner:2012cb, Boehm:2014vja, Kumar:2018yhh, Escudero:2018thh}. 
Meanwhile, current bounds on DM mass are often relying on the assumption that the abundance of DM is acquired through the thermal contact with the SM particles \cite{griest1990unitarity,Boyarsky:2008ju}. If DM transfers its entropy to either the electrons and photons or to the neutrinos \cite{Boehm:2013jpa},
one can set a lower limit on $m_\chi$ of $\mathcal{O}$(MeV) thorough its impact on the effective number of \minew{neutrinos} ($N_{\rm{eff}}$) \cite{Boehm:2012gr,Boehm:2013jpa,Ho:2012ug,Steigman:2013yua,Nollett:2014lwa,Steigman:2014uqa,Serpico:2004nm}. Nevertheless, thermal DM can be as light as a few keV
as long as dark matter thermalizes with the Standard Model below the temperature of neutrino-photon decoupling \cite{Berlin:2017ftj}. 

In this work, we focus on the interaction between DM and photons ($\gamma$) and purpose an alternative method to set limit on $m_\chi$. Note that this method is applicable to the interaction of DM with other standard model particles like
baryons or neutrinos.
By assuming elastic scattering between DM and photons, previous works \cite{Wilkinson:2013kia,Ali-Haimoud:2015pwa,Stadler:2018jin, Becker:2020hzj} have derived a upper bounds on the DM-photon($\gamma$) scattering cross sections ($\sigma_{\chi-\gamma}$) to mass ratio, e.g. $\sigma_{\chi-\gamma}\leq8\times 10^{-31}(m_\chi/\rm{GeV})$ at $68\%$ C.L. 
This ratio can be translate to the DM-$\gamma$ decoupling scale ($a_{\rm{dec}}$) \cite{Becker:2020hzj}.
Note that the temperature of DM traces the temperature of photons ($T_\chi\propto a^{-1}$) prior to DM-$\gamma$ decoupling,
and dilute as: $T_\chi\propto a^{-2}$ after $a_{\rm{dec}}$. 
At the beginning, we assume that DM enters equilibrium with the SM at early times
and acquires a large thermal abundance as relativistic species.
If DM cools naturally with the Hubble flow, 
the mass of DM will be too small ($m_\chi\lesssim \rm{1eV}$) to be compatible with current observations \cite{Hochberg:2015fth,Hochberg:2015pha,Schutz:2016tid,Knapen:2016cue}.
Thus, we are left with the possibility that the elastic scattering is established when DM has already become nonrelativistic\footnote{It is possible
that DM coupled to photons at early time as relativistic species. In this case one has $m_\chi\gtrsim\mathcal{O}$(MeV), because DM shall transfer most of its entropy to SM particles prior to big ban nucleosynthesis (BBN) as it becomes nonrelativist \cite{Boehm:2012gr,Boehm:2013jpa,Ho:2012ug,Steigman:2013yua}.}. This requires the DM mass exceeding its kinetic energy when entering equilibrium with photons, i.e., $m_\chi \gg T_{\chi}(a_{\rm{eqm}})\equiv T_{\gamma}(a_{\rm{eqm}})$.
Even so, the energy and pressure density of DM can still contribute non-negligibly at the background level. This motivates us to consider the equation of state (EoS) of DM.
Such a picture can be subscribed to the scenarios of generalized dark matter (so we denote as GDM hereinbelow) which is first proposed by Ref. \cite{Hu:1998kj} (see \cite{Thomas:2016iav,Kopp:2018zxp} for a detailed discussion).
The EoS of DM is dependent on both the DM mass ($m_\chi$) and $\sigma_{\chi-\gamma}$.  
We include an extra parameter $m_\chi$ to study the EoS of DM and further its distinctive imprints on the CMB angular power spectrum.
These imprints allows us to constrain $m_\chi$ can independent of $\sigma_{\chi-\gamma}$, and thus one can set a lower limit on $m_\chi$ with \emph{Planck} data alone.
In the study of cosmic concordance problem, we find that the thermalization of DM has negligible impact on the sound horizon at last scattering $r_s(z_*)$ as well as the late expansion history, and thus does not solve the $H_0$ tension.
While the most prominent feature of this scenario is the suppression in small-scale modes, which lowers the inferred value of $\sigma_{8}$ and $S_8 \equiv \sigma_8(\Omega_m/0.3)^{0.5}$.

To study the transitional epoch of the DM-$\gamma$ decoupling, we have also investigated the background bulk viscous pressure $\Pi$ \cite{Velten:2021cqj,Zimdahl:1996fj}.
We find that the (negative) bulk viscous pressure can be viewed as a correction to the overestimated DM pressure when the two fluids are approximated by a single coupled fluid. Meanwhile, the impact of bulk viscous pressure on the background evolution of the universe is negligible when compared with the influence of the EoS of DM. 
The outline of this paper is as follows: in Sec. \ref{sec:theory} we take a brief review of the basic equations associated with DM-$\gamma$ interaction. 
In Sec. \ref{sec:phenomenology}, we introduce the basic features of the GDM scenario at both the background and perturbation levels. 
In Sec. \ref{sec:data}, we describe our numerical implementation of the GDM model and the datasets used in our analysis. 
The numerical results of the Markov chain Monte Carlo (MCMC) analysis are presented in
Sec. \ref{sec:discussion}. The discussion and conclusions are presented in Sec. \ref{sec:conclusions}.\\
\section{THEORY}\label{sec:theory}
In this section we present the basic ingredients of the
GDM model.
\subsection{Background}\label{sec:background}
At early times, the elastic scattering between dark matter and photons keeps the two species in kinetic equilibrium.
When the momentum exchange rate between the two species falls off the Hubble expansion rate, DM is assumed to decouple from photon-bath after which its temperature $T_{\chi}$ evolves as $a^{-2}$. The evolution of DM temperature is expressed as:
\begin{eqnarray}
\dot{T}_{\chi} = -2\mathcal{H}T_{\chi} + \frac{8\rho_{\gamma}}{3\rho_{\chi}}\dot{\mu}(T_{\gamma}- T_{\chi}),
\end{eqnarray}
where $\cdot \equiv \frac{d}{d\tau}$, $\mathcal{H}\equiv aH$ is conformal Hubble rate and $\dot{\mu}$ is the DM-$\gamma$ scattering rate defined as 
\begin{equation}
\dot{\mu}\equiv an_{\chi}\sigma_{\chi-\gamma}, 
\end{equation}
which is in analogous to the Thomson scattering rate $\dot{\kappa}\equiv an_{e}\sigma_{\rm{Th}}$, 
with $\sigma_{\chi-\gamma}$, $\sigma_{\rm{Th}}$ the DM-$\gamma$ scattering cross section and Thomson scattering cross section, respectively.
Following form previous works \cite{Wilkinson:2013kia,Stadler:2018jin}, we parameterize the effect of interaction by
dimensionless quantity
\begin{eqnarray}
	u_{\chi-\gamma}= \frac{\sigma_{\chi-\gamma}}{\sigma_{\rm{Th}}}\left(\frac{m_{\chi}}{100\:\rm{GeV}}\right)^{-1},
\end{eqnarray}
which is proportional to the ratio of $\dot{mu}$ and $\dot{\kappa}$. 
We denote as $a_{\rm{dec}}$ the scale at which the conformal DM-$\gamma$ momentum exchange rate $\frac{4\rho_{\gamma}}{3\rho_{\chi}}\dot{\mu}$ equals to the conformal Hubble rate $\mathcal{H}$, after
that DM decouple from photons. In the radiation-dominated era where $\mathcal{H}\propto1/a$, the decoupling scale
can be approximated as: 
\begin{eqnarray}
	a_{\rm{dec}} \approx 1.6 \times 10^{-3}\:u_{\rm{\chi-\gamma}}^{\frac{1}{2}}.\label{eq:a_dec}
\end{eqnarray}
CMB constraint in Ref. \cite{Stadler:2018jin} set an upper limit on the DM-$\gamma$ interaction ($u_{\chi-\gamma}\leq  2.25\times 10^{-4}$) and equivalently an upper bound on the decoupling scale, i.e, $a_{\rm{dec}} \lesssim 2.4 \times 10^{-5}$.

If the DM particles are relativistic at early times, 
the energy density and pressure are given by 
\begin{small}
	\begin{subequations}
		\begin{align}
		P_\chi(T_\chi(a,u_{\chi-\gamma})) &= \frac{g}{3}\int \frac{d^3p}{(2\pi)^3}\frac{p^2}{E_\chi}\frac{1}{e^{E_\chi/k_{\rm{B}}T_\chi(a,u_{\chi-\gamma})}\pm1}, \label{eq:relativistic_eos} \\
		\rho_\chi(T_\chi(a,u_{\chi-\gamma})) &=  g\int_{0}^{\infty} \frac{d^3p}{(2\pi)^3}E_\chi \frac{1}{e^{E_\chi/k_{\rm{B}}T_\chi(a,u_{\chi-\gamma})}\pm1},
		\end{align}
	\end{subequations}
\end{small}
with $E_\chi\equiv\sqrt{p^2+m_\chi^2}$, $g$ the internal degrees of freedom, 
and $\pm1$ corresponds to fermions/bosons, respectively.
The EoS of DM ($w_{\chi} = P_\chi/\rho_{\chi}$) determines the evolution of DM density:
\begin{small}
	\begin{eqnarray}
	\rho_\chi(a) = \rho_{\chi,0}\exp\left(3\int^{a_0}_{a}\frac{1+w_{\chi}\left(T_\chi(a',u_{\chi-\gamma}),m_\chi\right)}{a'}da'\right),\label{eq:rho_evolution}
	\end{eqnarray}
\end{small}
where $\rho_{\chi,0} \equiv \Omega_{\chi,0}\rho_{\rm{cr}}$ is the current density of DM, with $\rho_{\rm{cr}}$, $\Omega_{\chi,0}$
the critical density and DM density fraction, respectively. 
In relativistic approximation $(T_\chi\gg m_\chi)$, we have
\begin{eqnarray}
\rho_\chi(T_\chi) = \frac{g_*\pi^2}{30}T^{4}_{\chi} = \frac{g_*\pi^2}{30}T^{4}_{\gamma},\label{eq:rho_relativistic}
\end{eqnarray} 
with $g_*=g$ for bosons and $g_* = (7/8)g$ for fermions. 
From Eq. (\ref{eq:rho_relativistic}) and Eq. (\ref{eq:rho_evolution}), one can determine $m_{\chi}$ by equating $\rho_{\chi,0}$ with \emph{Planck} best-fit CDM density, i.e.,$\rho_{\chi,0}/\rho_{\rm{cr}}\equiv\Omega_{\chi,0} = \Omega_{c,0}$. 
Note that, the abundance of DM ($n_\chi\equiv \rho_{\chi}/m_\chi$) decrease with the increase of $a_{\rm{dec}}$, thus, an upper bound on $a_{\rm{dec}}$ can be translate to an upper limit on $m_\chi$ of $\mathcal{O}(\rm{eV})$.
This bound is much below the threshold of warm DM \cite{Viel:2013fqw}. Consequently, DM must be nonrelativistic when the DM-$\gamma$ elastic scattering being established. 

In the nonrelativistic approximation, the pressure and energy density of DM reads:
\begin{eqnarray}
P_\chi = n_{\chi}k_BT_{\chi}, \qquad \rho_{\chi} = n_{\chi}m_{\chi} +\frac{3}{2}n_{\chi}k_BT_{\chi},\label{eq:eos}
\end{eqnarray}
The Friedmann equations in this scenario are written as:
\begin{small}
\begin{subequations}
	\begin{align}
	H^2(z) = H_0^2\left[\Omega_\Lambda+ \Omega_{r,0}a^{-4}+ \Omega_{b,0}a^{-3}+ \Omega_{\chi}(a)\right], \\
	\dot{H}(z) = -\frac{3}{2}H_0^2\left[1-\Omega_\Lambda+\frac{1}{3}\Omega_{r,0}a^{-4}+w_\chi(a)\Omega_\chi(a)\right], 
	\end{align}
\end{subequations}
\end{small}
where $\Omega_{r,0},\Omega_{b,0}$ and $\Omega_\Lambda$ are the present time density parameters for radiation,
baryon and dark energy, respectively, while $\Omega_{\chi}(a)\equiv\rho_\chi(a)/\rho_{\rm{cr}}$ is the density parameter for 
DM.

\subsection{Bulk viscous}
\begin{figure}
		\begin{center}
	\includegraphics[scale = 0.24]{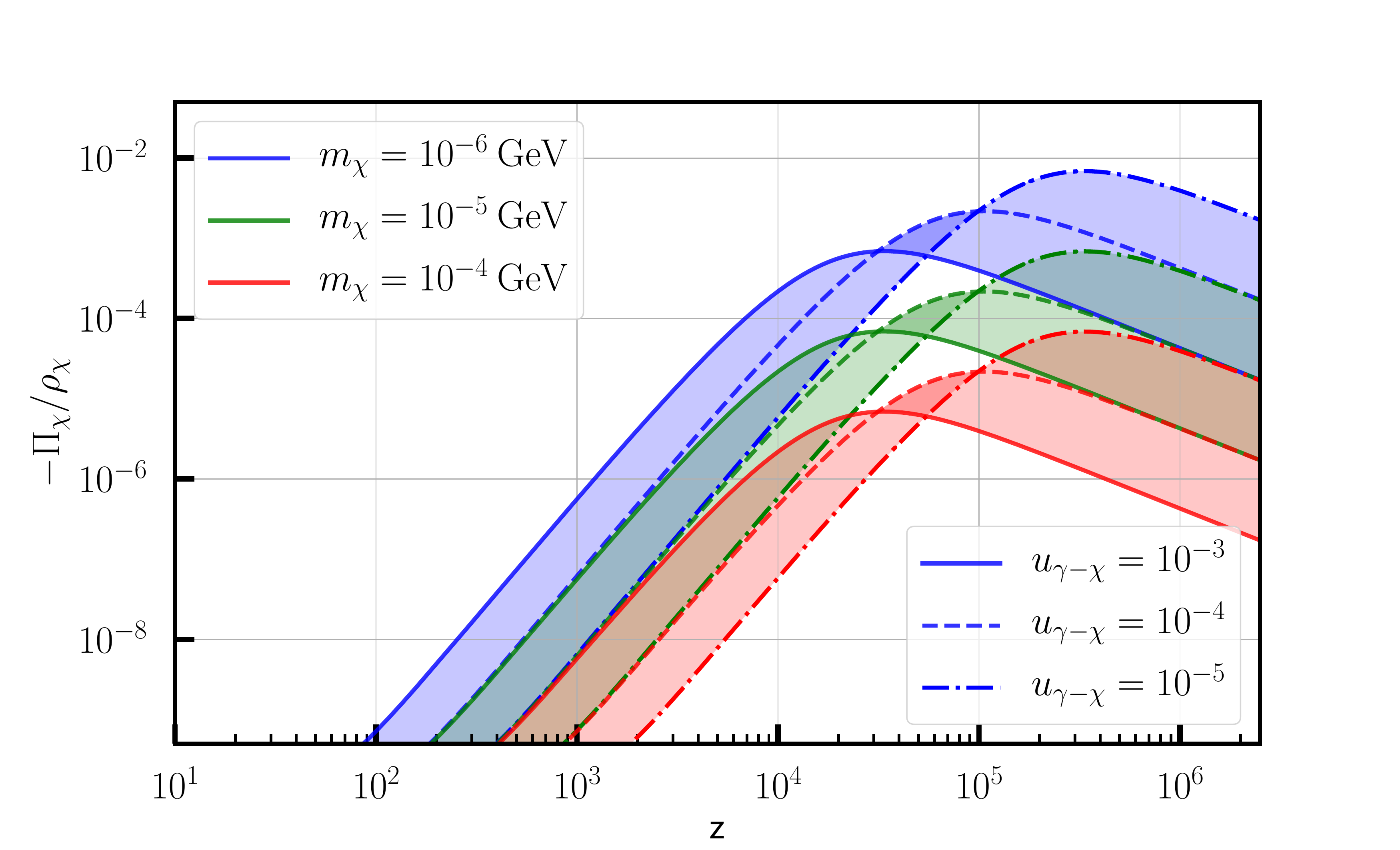}
	\includegraphics[scale = 0.24]{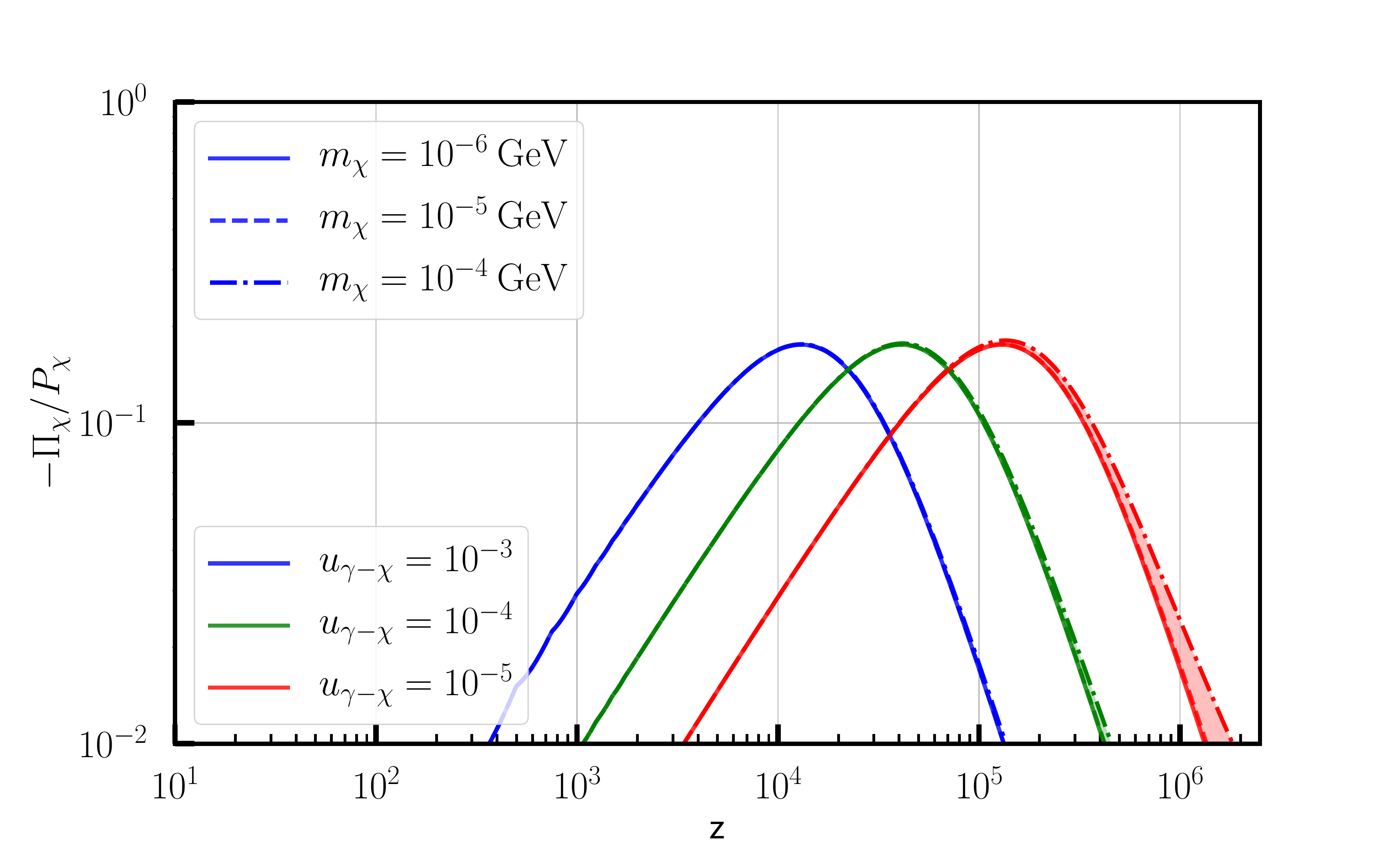}
		\end{center}
	\caption{
		     The ratio of negative bulk viscous pressure density to DM energy density $-\Pi_\chi/\rho_\chi$ (top panel);
		     The ratio of negative bulk viscous pressure density to DM pressure density $-\Pi_\chi/\rho_\chi$ (bottom panel).
	}\label{fig:rho_Pi}
\end{figure}

Works in Ref. \cite{Velten:2021cqj} have studied the emerging bulk viscous pressure $\Pi$, which acts as
an extra background effect during the DM-$\gamma$ transitional epoch.
We show in Appendix \ref{sec:viscous_append} that the bulk viscous pressure is a secondary effect of the thermodynamics of DM
at the background level, which can be viewed as a correction to 
overestimated DM pressure $P_\chi(T)$ when the two fluids are approximated by a single coupled fluid.
As is shown in the bottom panel of Fig. \ref{fig:rho_Pi}, the peak value of bulk viscous pressure $\Pi$ is of the order of $10^{-1}P_\chi$. Thus, the impact of bulk viscous pressure $\Pi$ on the cosmological background is negligible 
compared with the pressure density of GDM.  

\subsection{Perturbations}
Throughout the paper, we work in Newtonian gauge as is commonly used in numerical implementation, and our notation is followed from Ref. \cite{Ma:1995ey}.
We are interested only in scalar perturbations which are described by two potentials $\phi$ and $\psi$ and the line element is
\begin{equation}
ds^2 = a^2(\tau) \left[-\left(1+2\psi\right)d\tau^2 + \left(1-2\phi\right)dx_i dx^i\right]\,,
\end{equation}
where $\tau$ is the conformal time. We denote derivatives with respect to $\tau$ by a dot.
The pressure and energy density can be split into homogeneous background and small fluctuations, i.e., 
$P = \bar{P}+\delta P$ and $\rho = \bar{\rho}+\delta \rho$.
In conformal Newton gauge, the continuity and Euler equations for the fluctuations of fluid read \cite{Ma:1995ey}
\begin{subequations}
	\begin{align}
	\dot{\delta} &= -3\mathcal{H}\left(\frac{\delta P}{\delta \rho} - w\right)\delta - (1+w)\left(\theta-3\dot{\phi}\right)\,,
	\\
	\dot{\theta} &= -\mathcal{H}(1-3w)\theta - \frac{\dot{w}}{1+w}\theta + \frac{\delta P/\delta\rho}{1+w}\,k^2\delta- k^2\sigma + k^2\psi\,,
	\end{align}
	\label{eq: em-conservation-single}
\end{subequations}
with $k$ the comoving wavenumber, $\sigma$ the shear stress,
$\delta\equiv\delta \rho/\rho$ the density perturbation and $\theta$ the velocity dispersion.
The total entropy perturbation is defined as
\begin{eqnarray}
	\mathcal{S}\equiv \mathcal{H}\left(\frac{\delta P}{\dot{P}-\frac{\delta \rho}{\dot{\rho}}}\right)
\end{eqnarray}
In this work, we consider only isentropic perturbations, i.e., S=0. In this case, the adiabatic sound speed is written as
\begin{eqnarray}
c^2_{s} = \frac{\delta P}{\delta \rho} = \frac{\dot{P}}{\dot{\rho}}.\label{eq:cs}
\end{eqnarray}
Given the EoS (see Eq. (\ref{eq:eos})), we get immediately the sound speed of DM:
\begin{eqnarray}
c^2_{s(\chi)} =\frac{\delta P}{\delta \rho} =  \frac{\dot{P_\chi}}{\dot{\rho_\chi}} \approx 
\frac{k_BT_\chi}{m_\chi}\left(1- \frac{1}{3}\frac{d\ln T_\chi}{d\ln a}\right).\label{eq:sound_speed}
\end{eqnarray} 
The sound speed and EoS of DM are of the same order, meanwhile, the relationship between the two is written as
\begin{eqnarray}
\dot{w}_{\chi} = 3(1+w_\chi)(w_\chi-c^2_{s(\chi)})\mathcal{H}.\label{eq:w_dot} 
\end{eqnarray} 
Substituting Eqs. (\ref{eq:sound_speed}), (\ref{eq:w_dot}) into Eq. (\ref{eq: em-conservation-single}) we get 
the equations that collisionless DM obeys in a spatially flat Universe. 
With the addtion of DM-$\gamma$ collisional
term, these equations become
\begin{subequations}
	\begin{align}
	\dot{\delta}_\chi &=  - (1+w_\chi) \left(\theta_\chi-3{\dot{\phi}}\right)
	- 3\mathcal{H} \left(c^2_{s(\chi)} - w_\chi\right)\delta_\chi\,,
	\label{eq: DM-delta}\\
	\dot{\theta}_\chi &=  -\mathcal{H} (1-3c^2_{s(\chi)})\theta_\chi + \frac{c^2_{s(\chi)} k^2}{1+w_\chi}\delta_\chi + k^2 \psi -R\dot{\mu}(\theta_\chi -\theta_{\gamma})\,,
	\label{eq: DM-theta}
	\end{align}
\end{subequations}
	where $R\equiv 4\rho_{\gamma}/3\rho_{\chi}$, $\theta_{\gamma}$ and $\theta_{\chi}$ are velocity dispersion
	for photon and DM, respectively. The last term in Eq. (\ref{eq: DM-theta}) is the collisional term, which is nonzero only in the presence of DM-$\gamma$ interactions.
Note that we have not taken into consider the presence of shear stress $\sigma_{\chi}$.
The introduction of DM-$\gamma$ interaction
changes both the Boltzmann 
equations for DM and photons. 
Assuming that the DM-$\gamma$ scattering amplitude has the same angular polarization dependence as the Thomson elastic scattering cross section,
we can add the DM-$\gamma$ collision term in the equations governing the evolution of the photon fluid additional to Thomson interaction.
Thus, the decomposed Boltzmann equation for photons augmented by a new interaction term (see detailed derivations 
and higher orders of the Legendre polynomial decomposition of the energy distribution for photons in Ref. \cite{Stadler:2018jin}), written as:
\begin{small}
\begin{subequations}
	\begin{align}
	\dot{\delta}_\gamma &= -\frac{4}{3}\theta_\gamma + 4\dot{\phi}\,,
	\label{eq: boltzmann-h-cg-F0}\\
	\dot{\theta}_\gamma &= k^2\left(\frac{1}{4}\delta_\gamma-\sigma_\gamma\right) + k^2\psi + \dot{\kappa}\left(\theta_b-\theta_\gamma\right) + \dot{\mu}\left(\theta_\chi-\theta_\gamma\right)\,,
	\end{align}
	\label{eq: boltzmann-gdm-photons}
\end{subequations}
\end{small}
where $\dot{\kappa}$ is the Thomson scattering rate defined as $\dot{\kappa} = an_e\sigma_{\rm{Th}}$, with $n_e$ the electron
number density and $\sigma_{\rm{Th}}$ the Thomson scattering cross section.

The initial conditions for the isentropic perturbations in the conformal Newtonian gauge are written as:
\begin{eqnarray}
\label{super2}
	&&\delta_\gamma = -{40 C\over 15+4R_\nu} = -2\psi \,,\qquad
	  \delta_{\chi} = {3\over 4}\delta_\gamma \,, \nonumber\\
	&&\theta_\gamma=\theta_{\chi}={10 C\over
		15+4R_\nu} (k^2 \tau)=\frac{1}{2}(k^2\tau)\psi \,, \\
	&&\psi = {20 C\over 15+4R_\nu} \,,\qquad
	\phi = \left( 1+{2\over 5}R_\nu \right)\psi \,.\nonumber
\end{eqnarray}
Where the functions $\psi,\phi$ represent metric perturbations in the conformal Newtonian gauge, and $C$ is arbitrary dimensionless constants.
Note that the neutrino energy fraction $R_\nu$ is nonzero only in the presence of neutrinos.
See detailed derivations and discussions in Ref. \cite{Ma:1995ey}.
Note that these initial conditions are valid since the DM are nonrelativistic when all modes of interest enter the comoving horizon. 

\begin{figure}
	\begin{center}
		\includegraphics[scale = 0.26]{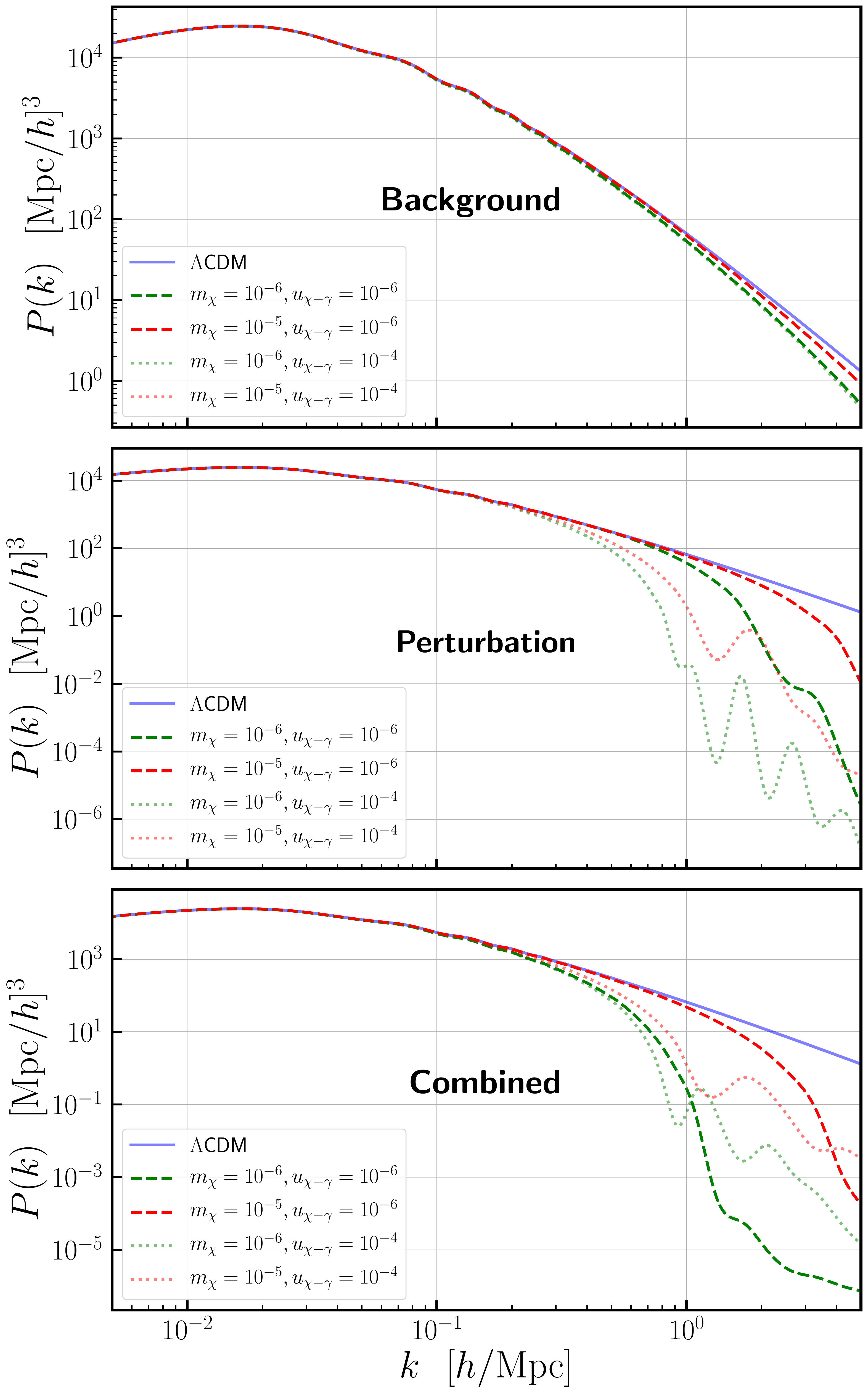}
	\end{center}
	\caption{The effect of  $m_\chi[\rm{GeV}]$ on matter power spectrum at the background level (top panel), 
		perturbation level (middle panel) and the combined results are shown in the bottom panel	
		The dashed line correspond to the spectrums with $u_{\chi-\gamma} = 10^{-6}$, and 
		the dotted line correspond to the spectrums with $u_{\chi-\gamma} = 10^{-4}$. 
	}\label{fig:Pk_gdm}
\end{figure}

\begin{figure*}
	\begin{center}
		\includegraphics[scale = 0.26]{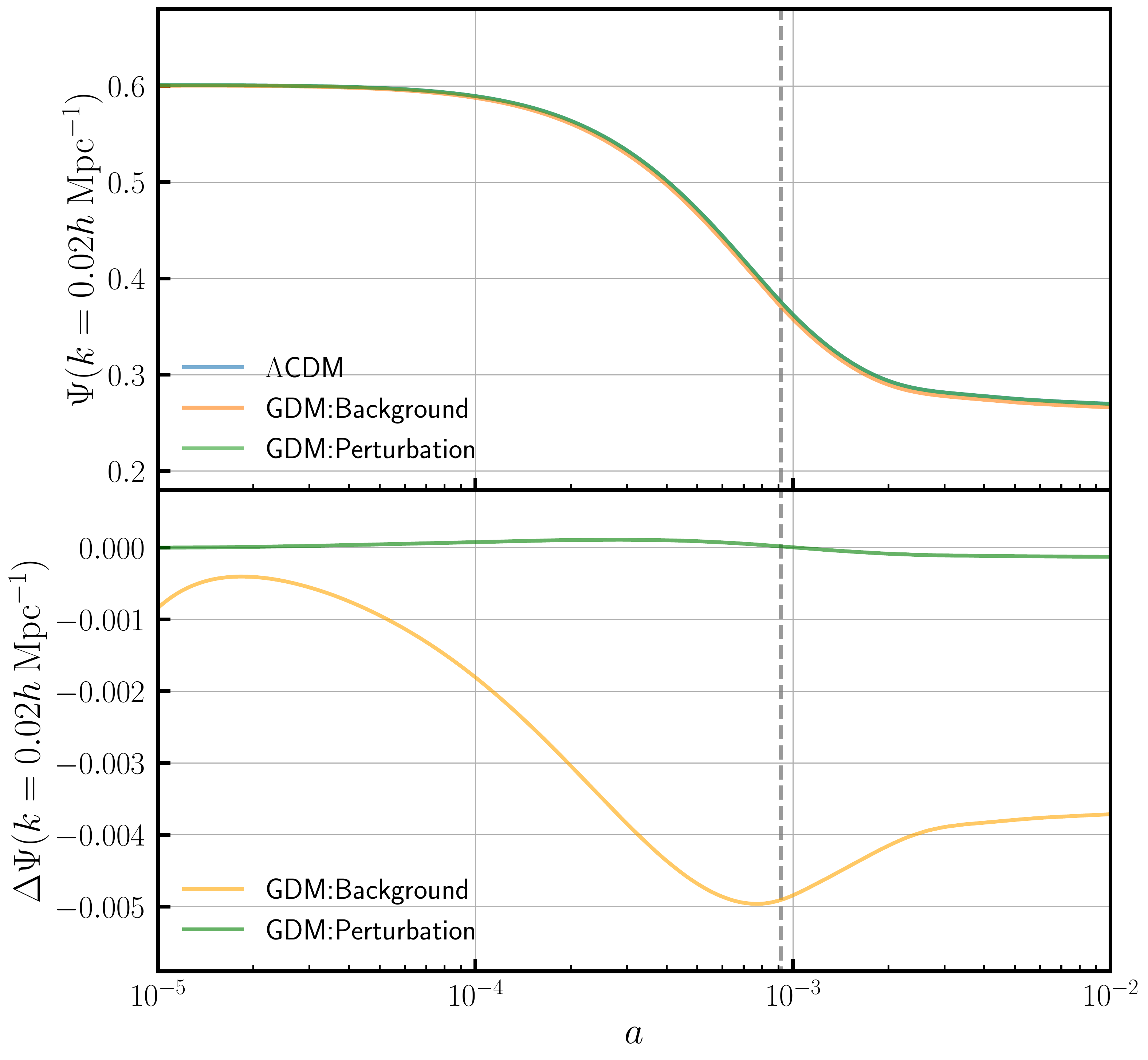}
		\includegraphics[scale = 0.26]{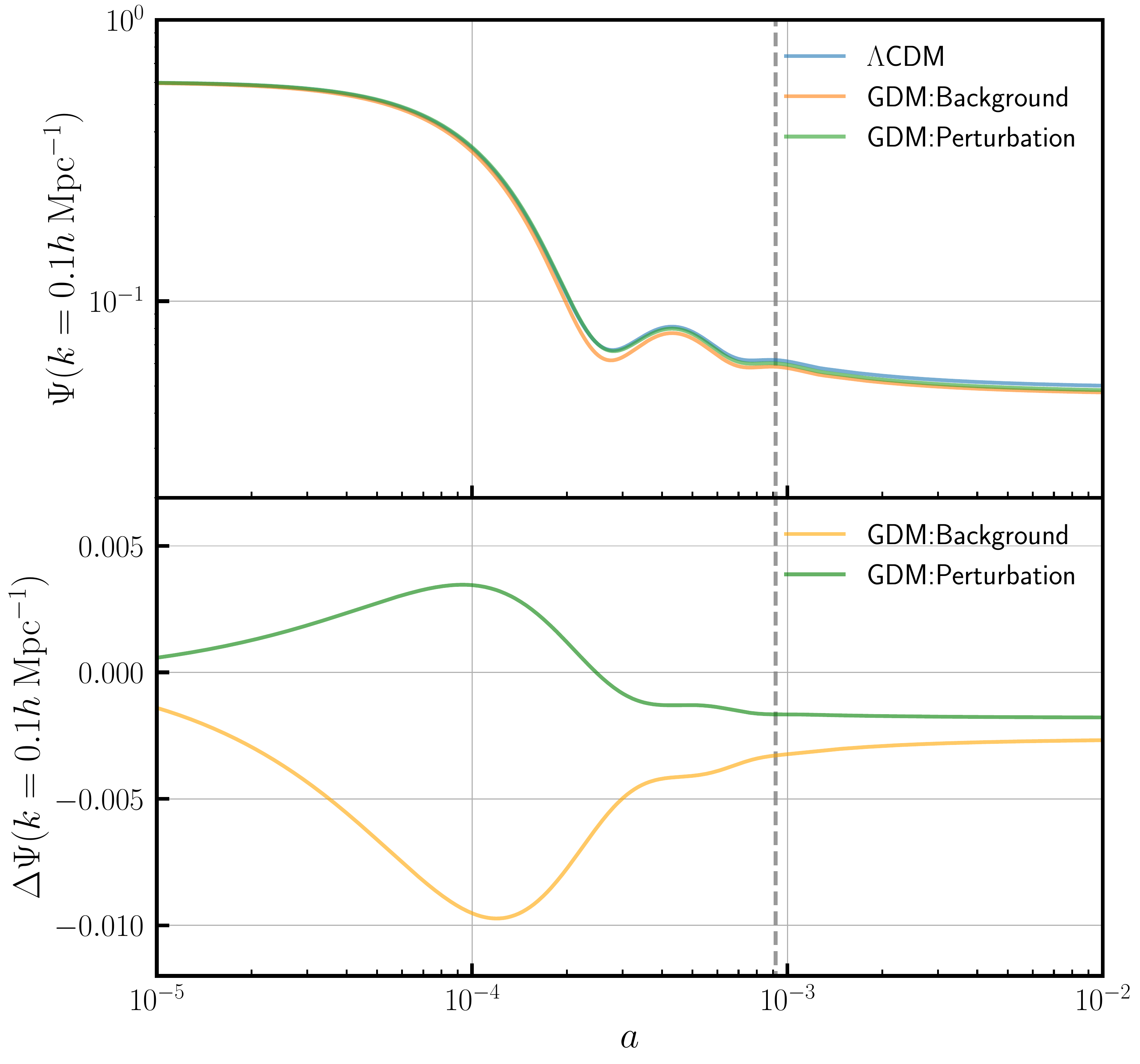}
	\end{center}
	\caption{The effect of GDM on the evolution of scalar potentials $\Psi\equiv\psi$ at $k = 0.02h\rm{Mpc}^{-1}$ and
		$k = 0.1h\rm{Mpc}^{-1}$, respectively. $\Delta\Psi\equiv \Psi^{\Lambda\rm{CDM}}-\Psi^{\rm{GDM}}$ is the relative change in the $\Psi$ with \emph{Planck} best-fit $\Lambda$CDM model as baseline. 
		We separate the effect into background and perturbation parts, while $m_\chi$ is 
		fixed at $10^{-7}$ GeV. 
	}\label{fig:psi}
\end{figure*}

\section{PHENOMENOLOGY}\label{sec:phenomenology}
In this section, we discuss the impact of DM-$\gamma$ interaction on cosmological observations at both background and perturbation levels. 
The effect of DM-$\gamma$ scattering on CMB and matter power spectrum (captured by the parameter $u_{\chi-\gamma}$)
have been elaborately discussed in previous works \cite{Wilkinson:2013kia,Stadler:2018jin}.
Whereas, in this work we focus on the influence of EoS of DM (captured by the parameter $m_{\chi}$).
As is discussed in Sec. \ref{sec:background},  DM decouples from DM-$\gamma$ plasma in radiation-dominated era.
Before the decoupling, the temperature of DM is equal to that of photons which evolves as $a^{-1}$, during this epoch, the EoS of DM is determined mainly by its masses rather than $u_{\chi-\gamma}$, which allow us to study characteristic imprints of $m_\chi$ on the CMB power spectra.  
At early times, the kinetic energy of DM particles is non-negligible and can be effectively be viewed as an ``extra'' energy density component (compared with the cold DM scenario) which alters the expansion history in the early universe. 
The DM-$\gamma$ scattering cross section is much smaller than the Compton scattering cross section, hence, the DM-$\gamma$ decoupling is much earlier than the last scattering. Moreover, the temperature of DM dissipates quadratically after the DM-$\gamma$ decoupling.
Consequently, the temperature of DM is sufficiently low that after the last scattering, which enables us to neglect the impact of DM pressure on the late-time growth faction $D(z)$ and growth function $f(z)\equiv d\ln D(z)/d\ln a$.


\begin{figure}
	\begin{center}
		\includegraphics[scale = 0.26]{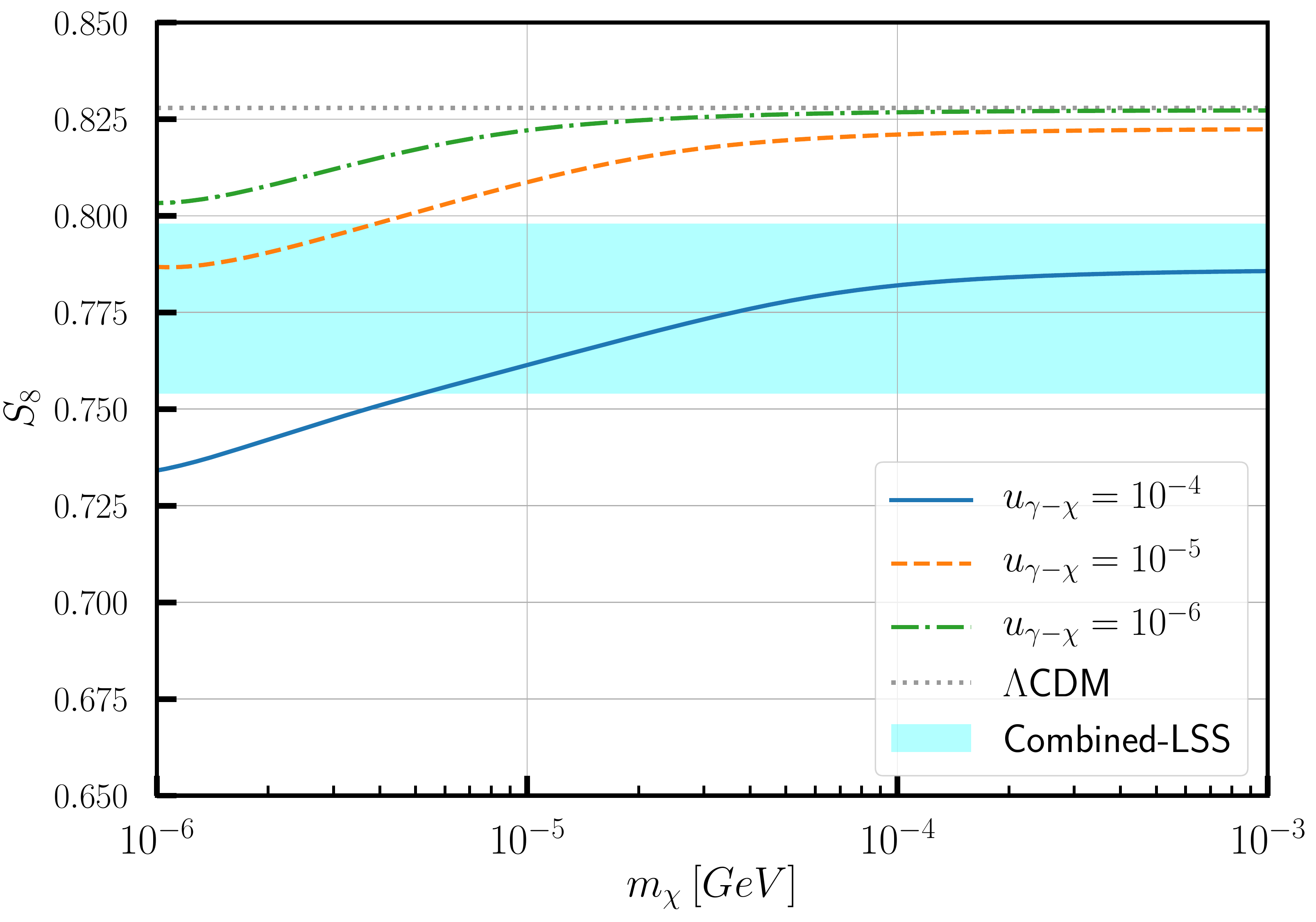}
		\includegraphics[scale = 0.26]{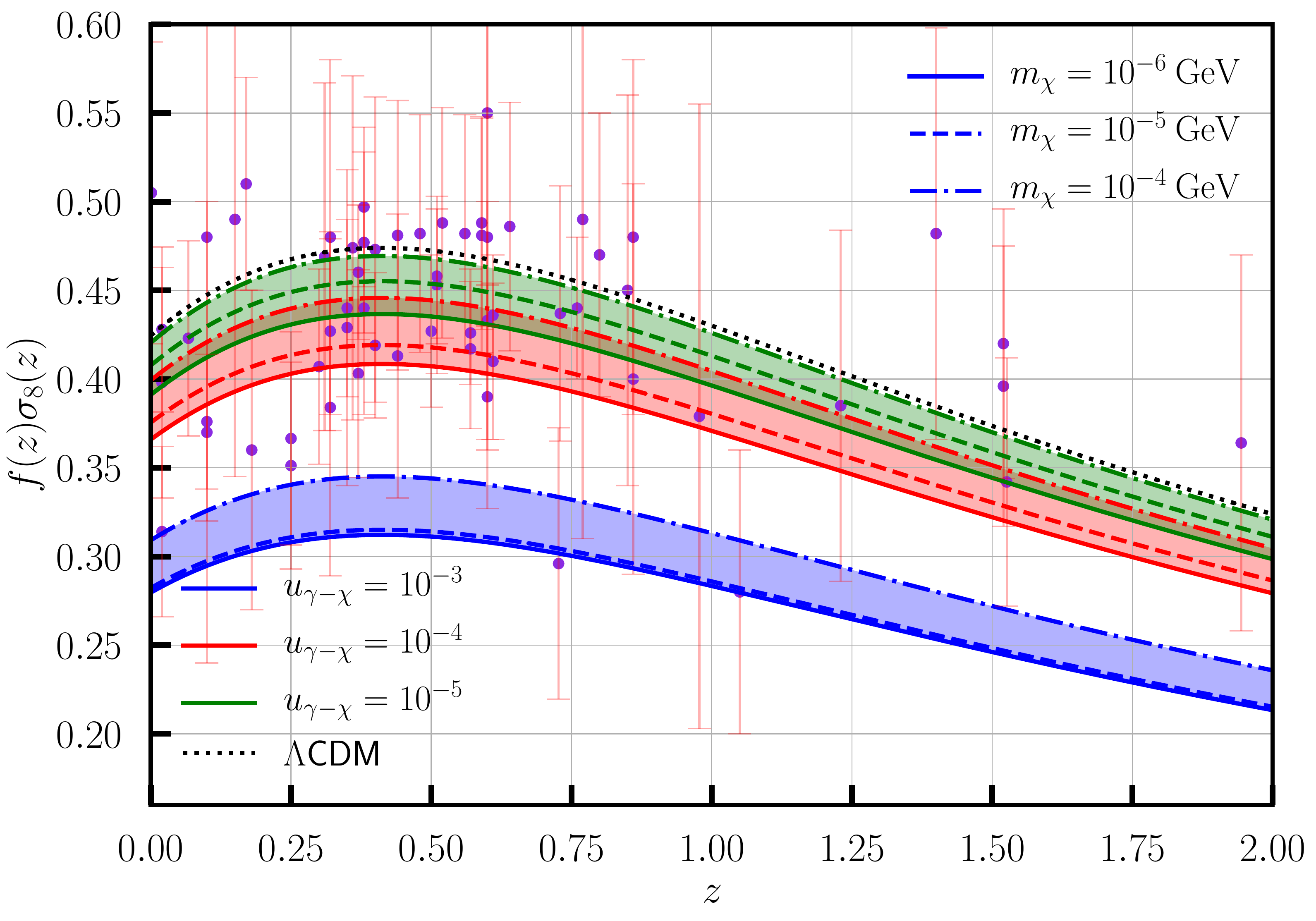}
	\end{center}
	\caption{The $S_8$ values calculated in the GDM model as a function of $m_\chi$ are shown in the top panel, and
			$f(z)\sigma_{8}(z)$ calculated in the GDM model are shown in the bottom panel.
			The horizontal band in the top panel corresponds to the DES-Y1
			constrains on $S_8$, i.e, $S_8 = 0.773^{+0.026}_{-0.020}$. The dotted line show the value calculated
			in the \emph{Planck} best-fit $\Lambda$CDM model. One should note that all 63 observational $f\sigma_{8,obs}(z)$ RSD data points collected by Ref. \cite{Kazantzidis:2018rnb} 
			are obtained assuming the fiducial $\Lambda$CDM cosmology.  Thus, the Alcock-Paczynski (AP) effect \cite{Alcock:1979mp} should be
			taken into account. In the present paper, we approxmate the AP effect as \cite{Macaulay:2013swa}: 
			$f\sigma_{8,\text{AP}}(z) \simeq \frac{H(z) D_A(z)}{H^{\rm{fid}}(z,\Omega_m) D_A^{\rm{fid}}(z,\Omega_m) } f\sigma_{8,\text{obs}}(z)$,
			where $H^{\rm{fid}}(z), D_A^{\rm{fid}}(z)$ are calculated based on the fiducial $\Lambda$CDM model, 
			while $H(z), D_A(z)$ are calculated in the GDM model.
	}\label{fig:s8_fs8}
\end{figure}


\begin{figure}
	\begin{center}
		\includegraphics[scale = 0.25]{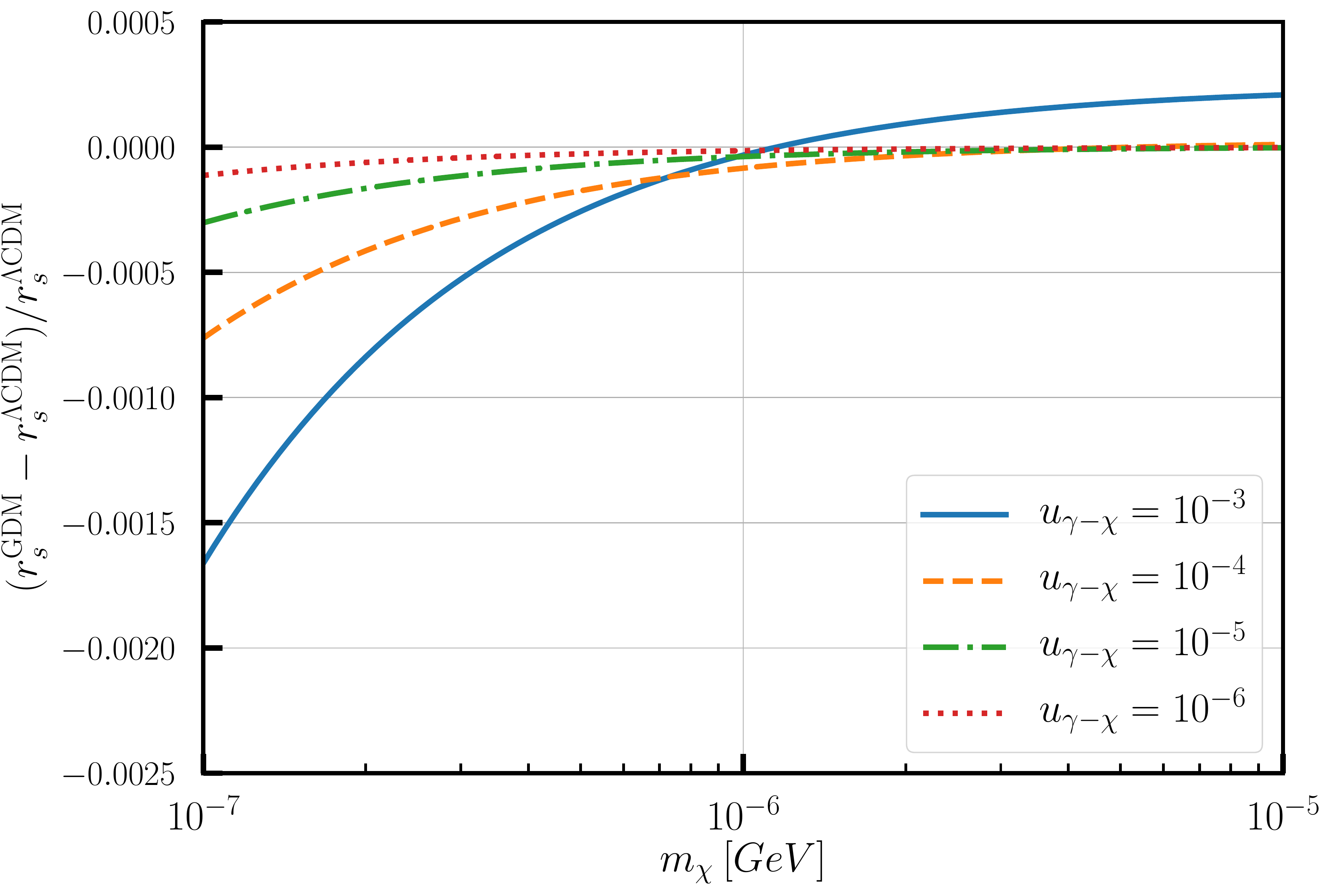}
	\end{center}
	\caption{The difference between the $r_s(z_*)$ calculated in the GDM model and in $\Lambda$CDM model, i.e., $(r_s^{\rm{GDM}}(z_*)-r_s^{\Lambda\rm{CDM}}(z_*))/r_s^{\Lambda\rm{CDM}}(z_*)$. 
	}\label{fig:delta_rs}
\end{figure}

\begin{figure}
	\begin{center}
		\includegraphics[scale = 0.26]{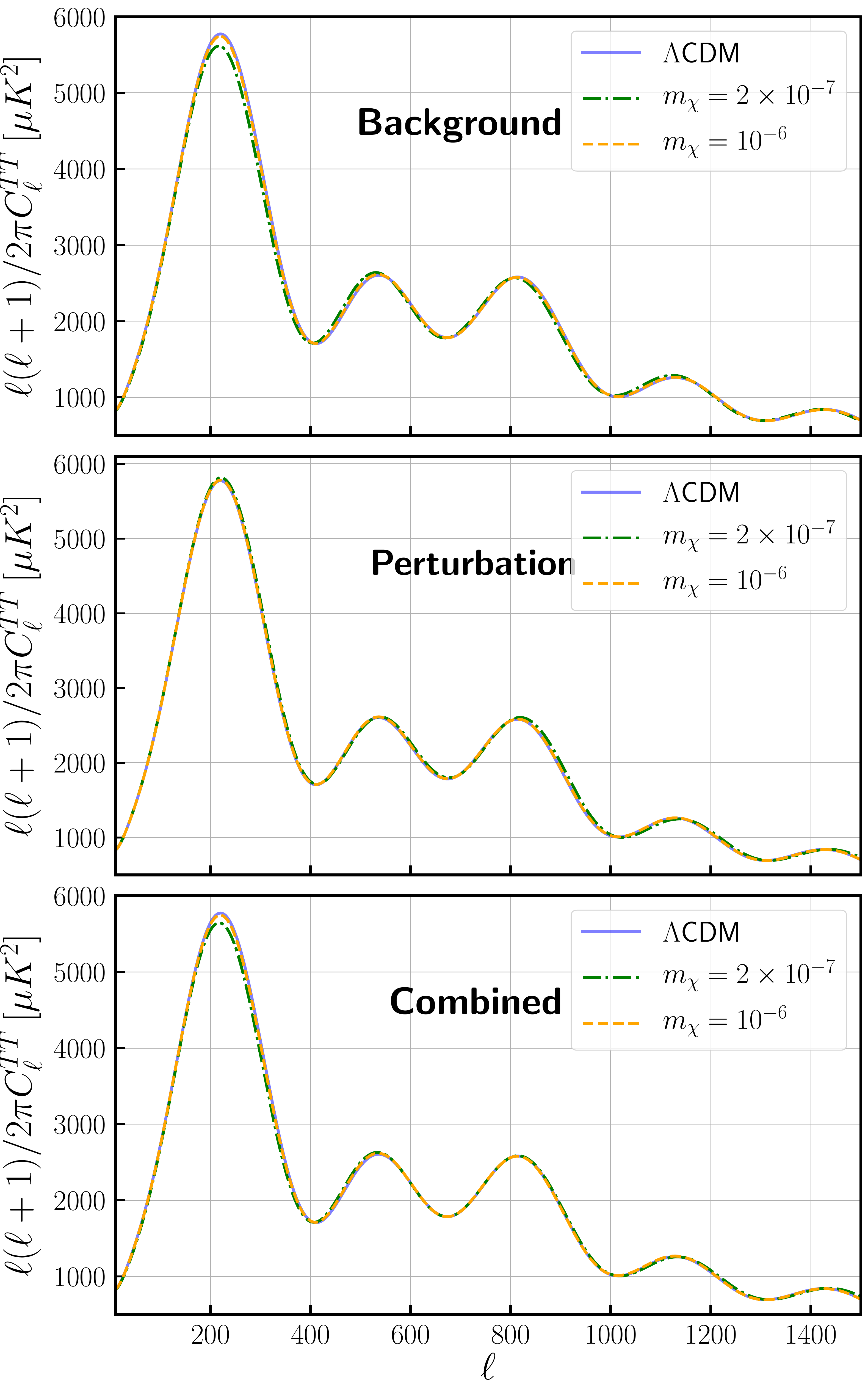}
	\end{center}
	\caption{The temperature angular power spectrum calculated in the GDM scenario with modifications to $\Lambda$CDM model
		separated into background part (top panel), perturbation part (middle panel) and their combinations (bottom panel). 
		Noting that $u_{\chi-\gamma}$ is fixed at $10^{-4}$.
	}\label{fig:cl_TT}
\end{figure}

\begin{figure}
	\begin{center}
		\includegraphics[scale = 0.26]{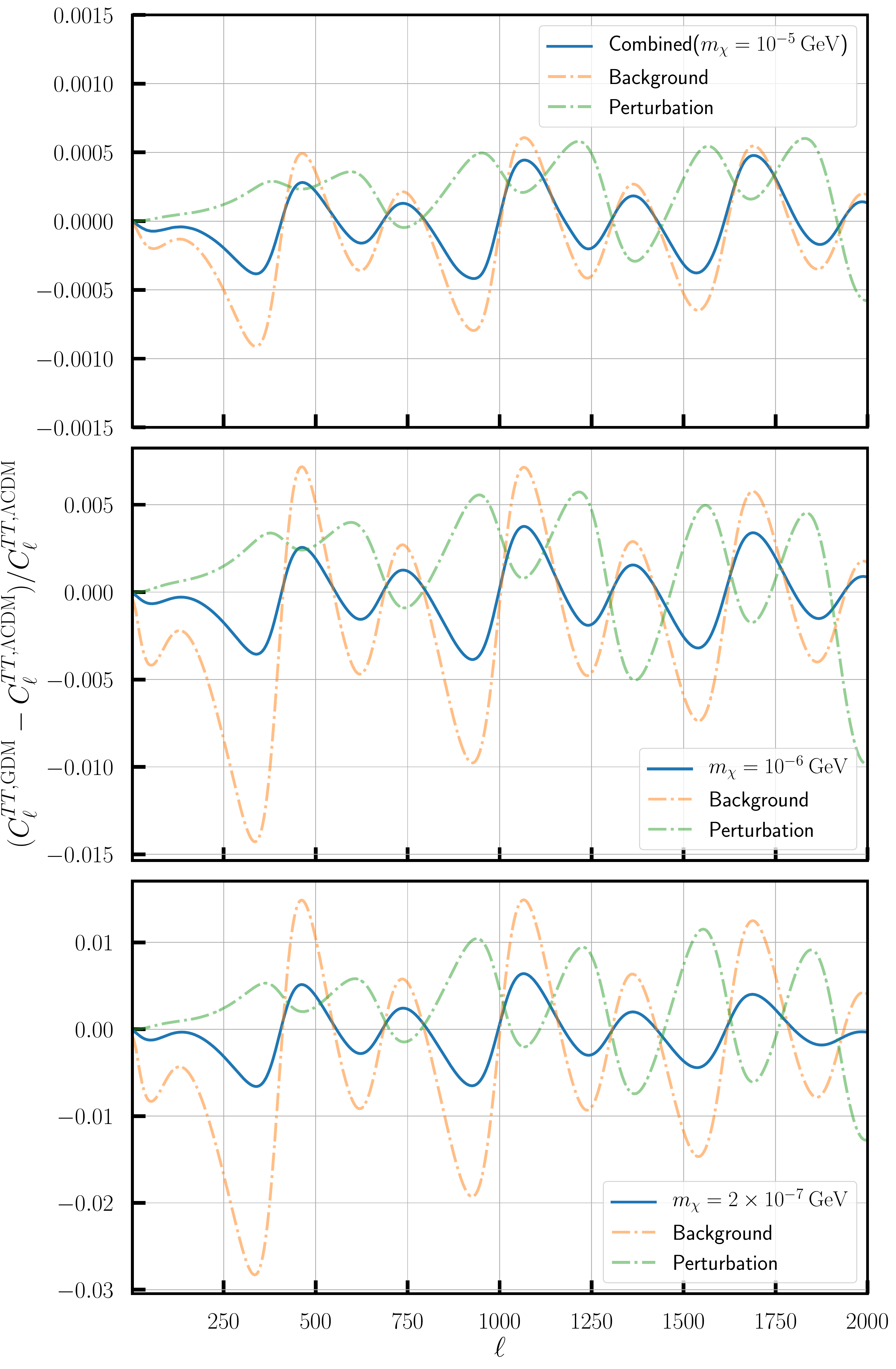}
	\end{center}
	\caption{The relative difference between the temperature angular power spectrum calculated in the GDM model and in $\Lambda$CDM model, $(C_{\ell}^{TT,\rm{GDM}}-C_{\ell}^{TT,\Lambda\rm{CDM}})/C_{\ell}^{TT,\Lambda\rm{CDM}}$. The modifications to $\Lambda$CDM model
		in the GDM scenario are separated into background part, perturbation part and their combinations. The relative difference 
		with $m_\chi$ fixed at different values are presented in separate panels.
	}\label{fig:Cl_diff}
\end{figure}

\subsection{Impact on the matter power spectra}\label{sec:matter_power}
To quantify the impact of EoS of DM on matter perturbations, we fix $u_{\chi-\gamma}$ at $10^{-6}$.
In this case, DM decouples from photons earlier than the modes of interest enter the comoving horizon, thus, the
small-scale suppression by collisional damping \cite{Wilkinson:2013kia} can be neglected.
The small-scale modes evolve within the linear regime at early times and start evolving nonlinearly 
during the matter domination. As is discussed previously,  the EoS of DM is negligible after last scattering \footnote{We show in Sec. \ref{sec:discussion} the \emph{Planck} constraints on DM masses and $u_{\chi-\gamma}$.
If we set $m_\chi$ to the lower limit  ($10^{-5}$ GeV) and $u_{\chi-\gamma}$ to the upper limit ($1.6\times10^{-4}$),
then the temperature and EoS of DM at last scattering reads: $T_\chi(z_*) \approx 10^{2}$K and $w_\chi(z_*) \approx 10^{-6}$.}, and thus has no impact on the evolution of perturbation in late (nonlinear) regime. Consequently, 
we only consider its influence on the linear matter power spectrum. 

To give a clear view, we separate the modifications to $\Lambda$CDM model into the ``background'' part, the ``perturbation'' part and their combinations.
By background, we mean that only the modifications to background quantities are considered, e.g., $\rho_\chi$
and $P_\chi$,
meanwhile, the equations governing the perturbations are identical with that of $\Lambda$CDM model, i.e., $\dot{\delta}_{\chi}= - \theta_\chi
+3\dot{\phi},\: \dot{\theta}_{\chi} = -\mathcal{H}\theta_{\chi} + k^2\psi$ \cite{Ma:1995ey}.
Accordingly, perturbation means that the background quantities are the same with that of the CDM scenario, while only the modifications to the continuity equation [Eq. (\ref{eq: DM-delta})], Euler equation [Eq. (\ref{eq: DM-theta})] and Boltzmann equations for photons [Eq. (\ref{eq: boltzmann-gdm-photons})] are considered. 
We obtained $u_{\chi-\gamma}<1.58\times10^{-4}$ and $m_\chi>0.87\times10^{-5}$ GeV by fitting with the Planck data 
(see sectiion \ref{sec:discussion}).
Accordingly, we choose $u_{\chi-\gamma}= 10^{-4}$ and $m_\chi = 10^{-5}$ GeV to illustrate the impact of 
DM-$\gamma$ scattering (captured by $u_{\chi-\gamma}$) and EoS of DM (captured by $m_\chi$). 
To isolate the impact of the parameter $m_\chi$, we choose $u_{\chi-\gamma}= 10^{-6}$ so that
its impact on matter power spectrum is negligible. As can be seen in the top panel of Fig. \ref{fig:Pk_gdm}, the most prominent feature is the suppression on small-scale modes which exacerbates
with the deceasing of $m_\chi$. This suppression can be attributed to the reduction in the gravitational potential $\Psi\equiv\psi$ (see Fig. \ref{fig:psi}).
For small-scale modes which enter the horizon during the radiation-dominated era, its potential starts to decay and oscillates after decaying. 
With the presence of ``extra'' pressure density $P_\chi$ (which is inversely correlated with $m_\chi$), the decay is even more rapid ($\dot{\Psi}$ \minew{decreases}).

At the perturbation level, we can see a similar suppression on the small-scale modes.
Compared with the background effect, the unset of the suppression is extrapolated to even smaller scales.
We have also included the matter power spectrum with $u_{\chi-\gamma}$ 
fixed at $10^{-4}$ as a comparison. 
As can be seen in the top panel of Fig. \ref{fig:Pk_gdm}, the increase in DM-$\gamma$ scattering with photons (form $u_{\chi-\gamma} = 10^{-6}$ to $u_{\chi-\gamma} = 10^{-4}$) have negligible influence on the spectra at the background level.
At perturbation level, owing to the delay of DM-$\gamma$ decoupling with the increase of $u_{\chi-\gamma}$ ($a_{\rm{dec}}\propto u_{\chi-\gamma}^{1/2}$) the suppressions on small-scale modes are aggravated .
Meanwhile, due to the increase in temperature of DM, one can see a more clear feature of acoustic oscillations in the matter spectrum (see Eq. (\ref{eq:cs}) and discussion in Ref. \cite{Stadler:2018jin}). 

The amplitude of spectrum at small scales is efficiently captured by the parameter $\sigma_8$, which gives the fluctuations of mass within the spheres of radius $8h^{-1}$ Mpc. 
In this light, we use $\delta\sigma_8/\sigma_8$  (or equivalently $\delta S_8/S_8$) 
to quantify the small scale suppression suppression. When DM-$\gamma$ scattering cross section is sufficiently small, i.e., $u_{\chi-\gamma}\leq 10^{-6}$,
the infferred value of $\sigma_8$ is reduced by $0.5\%$ for $m_\chi = 10$ keV and $3\%$ for $m_\chi = 1$ keV. In the presence of considerable DM-$\gamma$ scattering ($u_{\chi-\gamma}\leq 10^{-4}$), the infferred value of $\sigma_8$ is reduced by $5\%$ for
$m_\chi\gtrsim 1$ MeV ($w_\chi\approx 0$), while the reduction reach to $6\%$ for $m_\chi=10$ keV and $11\%$ for $m_\chi = 1$ keV.
We show in Fig. \ref{fig:s8_fs8} that the larger $u_{\chi-\gamma}$ the smaller inferred value of $S_8\equiv \sigma_{8}(\Omega_{m,0}/0.3)^{\frac{1}{2}}$, meanwhile, $S_8$ decreases with
the decrease of $m_\chi$.  The same is true for the combination $f(z)\sigma_{8}(z)\equiv f\sigma_{8}(z)$, as can be seen in Fig. \ref{fig:s8_fs8}.


\subsection{Impact on CMB spectra}
In this section, we discuss the impact of DM-$\gamma$ on the CMB anisotropy spectrum with the standard cosmological parameters kept fixed at \emph{Planck} 2018 best-fit values. 
We start our analysis by investigating the key observables.
As is shown in Fig. \ref{fig:rho_Pi}, GDM occupies a non-negligible fraction in the early universe, which shifts the matter-radiation equality scale and further affects the position and amplitude of the acoustic peaks in the angular power spectrum.
While a change in the expansion history of the universe $H(z)$ alters the comoving sound horizon $r_s(z_*)$ and angular diameter
distance $D_A(z_*)$ at last scattering surface, which determines the acoustic scale $\theta_s$ of the pecks.
However, our results suggest that within the scenario of GDM none of the above effects have considerable impacts on the shape of angular
power spectrum. 
As is discussed in the previous section, the EoS of DM is consistent with zero ($w_\chi\lesssim 10^{-6}$) after last scattering, and thus does not affect 
the late-time expansion history and $D_A(z*)$.
Although the increase of energy density for DM component ($\Delta \rho_{\rm{DM}} =\rho_{\chi} - \rho_{\rm{CDM}}$) does reduce the sound horizon $r_s(z_*)$, 
we show in Fig. \ref{fig:delta_rs} that the reduction is negligible ($\Delta r_s < 10^{-3}$) unless the mass of DM is sufficiently small ($m_\chi<10^{-7}$ GeV).
Hence, the spacings between the acoustic peaks are preserved.
Also, the shift in $z_{eq}$ is sufficiently small ($\Delta z_{eq} \sim 10^{-6}$), and thus should not be counted as the main contribution
to the relative differences.

To distinguish the impact of GDM on CMB angular power, the modifications to the $\Lambda$CDM model 
are separated into background and perturbation part (see Sec. \ref{sec:matter_power} for the details of the separation method).

\begin{enumerate}[(i)]
	\item In the background part, the main contribution for changes in $C_{\ell}$'s is the alteration in the monopole part of the perturbation, which free-stream into higher multipoles after last scattering, i.e.,
	\begin{eqnarray}
	\Theta_{\ell}^{\rm{Monopole}}(\eta_0) = [\Theta_0(\eta_*)+\Psi(\eta_*)]j_{\ell}(k(\eta_0-\eta_*)),
	\end{eqnarray}
	with $\eta_0$, $\eta_*$ the conformal time at present and last scattering, respectively; 
	and $\Theta_0\propto \delta_{\gamma}$. Note that the
	monopole is the dominant contribution to the $C_{\ell}$'s at small scales (see Ref.  \cite{Dodelson:1995es} for a
	detailed discussion).
	Apart from the effect on free-streaming monopole, the changes in the time evolution of $\Psi$ also affect the angular power spectrum through the early integrated Sachs-Wolfe (eISW) effect \cite{Sachs:1967er, Vagnozzi:2021gjh}. The remanent pressure of the GDM reduces the decay of gravitational potential around recombination scale $a_{\rm{rec}}\sim 10^{-3}$, i.e., $\Delta\dot{\Psi}>0$ (see Fig. \ref{fig:psi}), which leads to a depletion in the angular power spectrum on multipoles $50 < \ell < 500$ \cite{Lesgourgues:2014zoa}.
	As can be seen from the top panel of Fig. \ref{fig:cl_TT}, the most prominent feature is the damping of the first peak compared with the angular power spectrum obtained in the $\Lambda$CDM model.
	\item In the perturbation part, the trend of the evolution of $\Delta\Psi$ is opposite to the case of the background part.  
	Form the last panel of Fig. \ref{fig:psi}, one can notice a enhanced decay of the gravitational potential $\Psi$ at $a\gtrsim 10^{-4}$. This leads to a scale-dependent enhancement of the 
	eISW effect, i.e.,
	a increment in the angular spectrum around the first acoustic peak (see the middle panel of Fig. \ref{fig:cl_TT}).
\end{enumerate}
We show in Fig. \ref{fig:Cl_diff} the relative deviation of GDM scenario from the base-line \emph{Planck} 2018 $\Lambda$CDM model.
Comparing the relative deviation spectrum at background and perturbation level, one can notice phase displacement of the two patterns, while the combined results are the reductions of odd acoustic peaks and the boost in even peaks of the angular power spectrum (compared with $\Lambda$CDM model).
Meanwhile, the deviation increases with the decease of $m_\chi$, which indicates the CMB data alone is able to constrain DM masses.



\begin{table*}
	\renewcommand{\arraystretch}{1.4}
	\begin{center}
		\begin{tabular}{|llllll|}			
			\hline
			\hline			
			Dataset & \emph{Planck} & \emph{Planck}+lensing & \emph{Planck}+SH0ES & \emph{Planck}+$S_8$& \emph{Planck}+$S_8$+BAO \\
			\hline
			\hline
			$S_8$             & $0.819^{+0.025}_{-0.016}$ & $0.808^{+0.022}_{-0.014}  $ & $0.797^{+0.026}_{-0.017}$ &$0.785\pm 0.017 $ & $0.787\pm 0.016$ \\
			$\log_{10}(m_\chi/\rm{GeV})$         &   $> -5.06$               & $> -5.07$ & $-5.00$ & $>-5.06$& $>-5.07$ \\
			$u_{\chi-\gamma}$ &  $< 1.55 $                & $< 1.90$ & $<1.72 $ & $< 2.39\:(1.20^{+0.62}_{-0.80})$& $<2.20\:(1.15^{+0.54}_{-0.90})$ \\
			\hline
			$H_0$             & $67.59\pm 0.55    $       &$ 67.81^{+0.58}_{-0.50}  $ & $68.46\pm 0.57 $ &$67.99\pm 0.56$ & $68.01^{+0.44}_{-0.56}$\\
			$\Omega_bh^2$     &  $0.02243^{0.00016}_{-0.00015}    $   & $0.0225\pm 0.00014$ & $0.02259\pm 0.00014    $ &  $0.02252\pm 0.00014$& $0.0225\pm 0.00014 $\\
			\midrule
			$\Omega_ch^2$     & $0.1198\pm 0.0013 $       & $0.1193^{+0.0012}_{-0.0015}$ & $0.1179\pm 0.0013$ & $0.1193\pm 0.0015     $& $0.1192^{+0.0015}_{-0.0013}$\\
			\midrule
			$\log(10^{10} A_\mathrm{s})$ & $3.077^{+0.012}_{-0.013}$ & $3.0765\pm 0.0055$   & $3.0739\pm 0.0066 $ & $3.0769\pm 0.0063  $& $3.0760\pm 0.0067 $\\
			$n_\mathrm{s}   $ &$0.9667\pm 0.0041$& $0.9677\pm 0.0042 $ & $0.9717\pm 0.0041 $ &$0.9690\pm 0.0040 $& $0.9692^{+0.0035}_{-0.0043}$\\
			$\chi^2_\mathrm{tot} $ & $1014.6\pm 6.7$            & $1024.7\pm 4.1$     & $1033.3\pm 4.3$ &$1020.0\pm7.7$& $1027.1\pm 9.6  $\\
			\hline

		\end{tabular}
	\end{center}
	\caption{The the mean $\pm1\sigma$ constraints on the cosmological parameters and derived parameters in $\Lambda$CDM model (second column) and GDM model, as inferred from the combination of the \emph{Planck} 2018 low-$\ell$ TT+EE and high-$\ell$ TT+TE+EE power spectrum; the SH0ES prior on $H_0$; the $S_8$ prior derived from combined LSS datasets; the joint BOSS DR12 BAO and RSD data. Upper and lower bounds correspond to the $68\%$ C.L. interval. When only upper or lower limits are shown they correspond to $95\%$ C.L. limits (in this case the $68\%$ C.L. intervals are enclosed in brackets). }
	\label{tab:accumulate_constraints}
\end{table*}

\section{STATISTICAL METHODOLOGY AND DATASETS}\label{sec:data}
We implement the GDM scenario as modifications to
the publicly available Einstein-Boltzmann code \href{http://class-code.net/}{\texttt{CLASS}} \cite{Lesgourgues:2011re,Blas:2011rf} package.
The nonlinear matter power spectrum required by redshift-space distortion (RSD) likelihoods are computed using the ``HMcode'' \cite{Mead:2015yca,Mead:2016ybv,Mead:2020vgs} implemented in \texttt{CLASS}.
The MCMC analyses are performed 
using the publicly available code \href{https://cobaya.readthedocs.io}{\texttt{Cobaya}} \cite{Torrado:2020dgo} package
with a Gelman-Rubin \cite{10.1214/ss/1177011136} convergence
criterion $R-1 < 0.05$. The plots have been obtained
using the \href{https://getdist.readthedocs.io}{\texttt{GetDist}} \cite{Lewis:2019xzd} package.
The following datasets are considered in the MCMC analyses:
\subsection{CMB}
We employ the \emph{Planck} 2018 low-$\ell$ TT+EE and \emph{Planck} 2018 high-$\ell$ TT+TE+EE temperature and polarization power spectrum
\cite{Planck:2019nip}. To marginalize over nuisance parameters, we use the ``lite'' likelihoods. These datasets are referred to as ``\emph{Planck}''. We have also considered the \emph{Planck} 2018 lensing power spectrum \cite{Planck:2018lbu}.
\subsection{Hubble constant}
The most recent SH0ES measurement indicates that $H_0 = 73.2 \pm
1.3$ \cite{Riess:2020fzl}, indicating a tension at $\sim4\sigma$ with the \emph{Planck} \cite{Aghanim:2018eyx} value of $H_0= 67.4\pm 0.5$ assuming a minimal $\Lambda$CDM model. 
\subsection{LSS}
We consider the LSS datasets to probe the low-redshift universe, which include:
\begin{itemize}
	\item BAO \& RSD: SDSS BOSS DR12 \cite{Alam:2016hwk} measurements of the BAO signal and $f\sigma_8(z)$, at $z =0.38$, 0.51 and 0.61. We include the full covariance of the joint BOSS DR12 BAO and RSD data (denote as BAO). 
	\item Weak lensing:
	We consider the tomographic weak gravitational lensing analysis of 
	the Dark Energy Survey (DES-Y1) \cite{Abbott:2017wau}, the Kilo Degree Survey (KV450) \cite{Hildebrandt:2016iqg,Hildebrandt:2018yau} and Subaru Hyper Suprime-Cam (HSC) \cite{HSC:2018mrq}. 
	Following from Ref. \cite{Hill:2020osr}, the likelihoods for these datasets are not implemented directly. Instead, we approximately include their effect via priors on $S_8$. These experiments yield $S_8=0.770^{+0.018}_{-0.016}$ when combined with inverse-variance weights \cite{Hill:2020osr}.
\end{itemize}
we consider as base-line a 7-dimensional parameter space described by the following parameters: the Hubble constant $H_0$, 
the baryon density $\Omega_bh^2$, the dark matter energy density $\Omega_\chi h^2$, the scalar amplitude $A_s$,  the spectral index $n_s$,
the cross section to mass ratio $u_{\chi-\gamma}$ and the dark matter masses $m_\chi$. Uniform priors are assumed for 
all these parameters. 
In the base-line scenario we assume the mass of neutrinos $m_\nu = 0.06 eV$ and 
the effective number of neutrinos $N_{\rm{eff}} = 3.046$\footnote{The GDM is fully nonrelativistic when modes of interest enter the comoving horizon, and thus does not contribute to the effective number of neutrino species $N_{\rm{eff}}$. 
Meanwhile, allowing $N_{\rm{eff}}$ to vary does not have a significant effect on our conclusions.}
and the reionization optical depth $\tau_{\rm{reio}} = 0.054$. We set $u_{\chi-\gamma}>5\times10^{-7}$ ($a_{\rm{dec}}\gtrsim10^{-6}$) so that the DM-$\gamma$ decoupling have detectable imprints on the CMB. 
For $a_{\rm{dec}}<10^{-6}$, one has $m_\chi\gtrsim\mathcal{O}(10)$keV so that DM enters equilibrium with photons ($a_{\rm{eqm}}$) when it has already become nonrelativistic, i.e., $m_\chi\gg T(a_{\rm{eqm}}) >T(a_{\rm{dec}})$.

\begin{table*}
	\renewcommand{\arraystretch}{1.4}
	\begin{center}
		\begin{tabular}{|lllll|}			
			\hline
			\hline			
			Parameter & $\Lambda$CDM & $u_{\chi-\gamma}\geq 10^{-6}$ &  $u_{\chi-\gamma}\geq 10^{-5}$ & $u_{\chi-\gamma}\geq 10^{-4}$ \\
			\hline
			\hline
			$S_8$   &  $0.837^{+0.020}_{-0.016}   $ &  $0.819^{+0.025}_{-0.016}   $ & $0.817^{+0.024}_{-0.017} $ &$0.790^{+0.022}_{-0.017}$  \\
			$\log_{10}(m_\chi/\rm{GeV})$  &  $\dots$&$> -5.06$     & $> -5.05$ & $> -4.99$ \\
			$u_{\chi-\gamma}$  &  $\dots$ &  $< 1.55 $ & $< 1.59$ & $< 2.04$ \\
			\hline
			$H_0$  &  $67.65^{+0.60}_{-0.67}    $ & $67.59\pm 0.55  $ & $67.55\pm 0.58$ &$67.50^{+0.53}_{-0.67}$ \\
			$\Omega_bh^2$& $0.02241\pm 0.00017    $ & $0.02243\pm 0.00016$ & $0.02243\pm 0.00015 $ &$0.02246\pm 0.00016 $\\
			\midrule
			$\Omega_ch^2$ & $0.1198\pm 0.0013 $ & $0.1198\pm 0.0013$ & $0.1199\pm 0.0014 $ &$0.1205^{+0.0014}_{-0.0012}$\\
			\midrule
			$\log(10^{10} A_\mathrm{s})$ & $3.0750^{+0.0099}_{-0.027}$ & $3.077^{+0.012}_{-0.013}$ & $3.0779^{+0.0068}_{-0.0060}$ &$3.0782^{+0.0080}_{-0.0058}$ \\
			$n_\mathrm{s}   $ & $0.9663^{+0.0047}_{-0.0041}$ & $0.9667\pm 0.0041$ & $0.9663\pm 0.0042 $ & $0.9667\pm 0.0042$ \\
			$\chi^2_\emph{planck} $ & $1016\pm 23 $ & $1014.6\pm 6.7$ & $1015.2\pm 6.7$ &$1018.2\pm 4.1   $\\
			\hline

		\end{tabular}
	\end{center}
	\caption{The the mean $\pm1\sigma$ constraints on the cosmological parameters and derived parameters in the GDM scenario with different priors on $u_{\chi-\gamma}$. Upper and lower bounds correspond to the $68\%$ C.L. interval. When only upper or lower limits are shown they correspond to $95\%$ C.L. limits. }
	\label{tab:prior_constraints}
\end{table*}

\section{RESULTS AND DISCUSSIONS}\label{sec:discussion}

In this section, we fit the GDM model to different combinations of 
the above datasets and discuss the obtained results.
In Table \ref{tab:accumulate_constraints} we report the
constraints at $68\%$ C.L. on the standard cosmological parameters and some key derived quantities
for several datasets combinations, when only upper (lower) limits are shown they
correspond to $95\%$ C.L. limits. 
The constraint obtained from the \emph{Planck} 2018 datasets
with different priors on $u_{\chi-\gamma}$ are given in Table \ref{tab:prior_constraints}.
The triangular plot with the 1D posterior distributions and the 2D contour plots for these
parameters are shown in Fig. \ref{fig:gdm_contour} and Fig. \ref{fig:gdm_prior}.

\subsection{Results based on Planck}
To compare with previous work, we start by analysing the \emph{Planck} 2018 datasets.
The results obtained from \emph{Planck} 2018 low-$\ell$ TT+EE and \emph{Planck} 2018 high-$\ell$ TT+TE+EE are
shown in the first column of Table \ref{tab:accumulate_constraints}.
We find a $95\%$ C.L. upper
limit on the DM-$\gamma$ scattering cross section
$u_{\chi-\gamma} < 1.55 \times 10^{-4}$. This limit is
comparable with the result $u_{\chi-\gamma} < 1.58 \times 10^{-4}$ at $95\%$ C.L. obtained from \emph{Planck} 2015 TTTEEE + lowTEB datasets \cite{Planck:2015bpv}, which are derived in Ref. \cite{Stadler:2018jin}.
The derived value of $S_8 = 0.808^{+0.022}_{-0.014}$ in the GDM model is smaller than 
that of $\Lambda$CDM model due to the effect of collisional damping discussed in Ref. \cite{Wilkinson:2013kia}.  
When added with \emph{Planck} lensing datasets, we find that the upper limits on $u_{\chi-\gamma}$ are slightly increased 
while the inferred value of $S_8$ is mildly reduced, i.e., $u_{\chi-\gamma} < 1.90 \times 10^{-4}$ and 
$S_8 = 0.808^{+0.022}_{-0.014}$, respectively. As can be noticed in Fig, a slight upward shift on $u_{\chi-\gamma}$ corresponds to a larger DM-$\gamma$ decoupling scale. Thus, the onset of collisional damping is extrapolated to larger scale which leads
to smaller inferred value of $S_8$.
Meanwhile, we find the same $95\%$ C.L. upper limit on the dark matter mass with and without lensing dataset, i.e., 
$m_{\chi} > 5.06 \times 10^{-4}$ GeV, suggesting that within the GDM scenario the smallest allowed dark matter mass is roughly 10 keV.

\subsection{Including SH0ES}
\begin{figure*}
	\includegraphics[scale = 0.38]{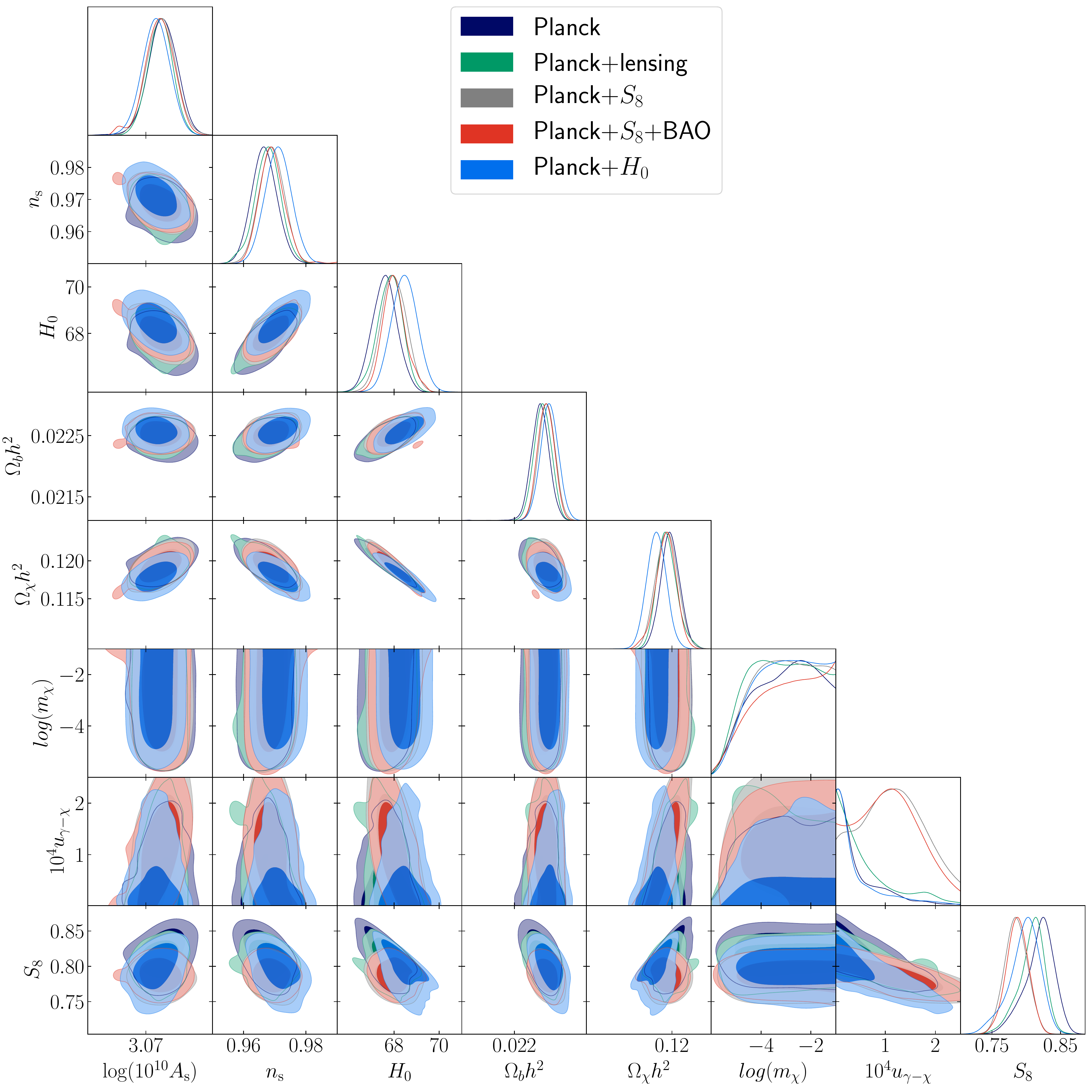}
	\caption{Constraints on cosmological parameters in the GDM scenario from 
		the \emph{Planck} 2018 low-$\ell$ TT+EE and high-$\ell$ TT+TE+EE power spectrum; \emph{Planck} 2018 lensing datasets;
		the SH0ES prior on $H_0$; the $S_8$ prior derived from combined LSS datasets; the joint BOSS DR12 BAO and RSD data. The
		contours show $1\sigma$ and $2\sigma$ posteriors for various dataset combinations. 
	 }\label{fig:gdm_contour}
\end{figure*}

To test the potential of the GDM scenario in relieving the $H_0$ tension, we conduct the joint analysis of SH0ES+\emph{Planck}
datasets, i.e., applying a Gaussian $H_0$ prior  in the fit of \emph{Planck} data.
In this analysis,  one would expect a significant shift of the best-fit $H_0$ value as long as the extra parameters in the GDM model  ($u_{\chi-\gamma}$, $m_\chi$) degenerate with $H_0$.
However, this is not the case, i.e., $H_0$ only shifts
very slightly and tensions with SH0ES remain at $2.5\sigma$. From the contours  in Fig. \ref{fig:gdm_contour} we do not see a clear degeneracy of $H_0$ with any given
parameters. Consequently, the GDM scenario does not relieve the $H_0$ tension.

\subsection{Including LSS datasets}
As is discussed in previous sections, there is a $1\sim3\sigma$ tension between the value
of $S_8$ predicted within $\Lambda$CDM with parameters fit from the CMB, and the value of $S_8$ from more direct measurements of LSS. 
The DM-$\gamma$ interaction suppresses the small-scale modes via both the reduction in free-streaming monopole ($\Theta_0(\eta_*)+\Psi(\eta_*)$) and diffusive damping.
Thus, it's natural to think of the GDM scenario as a viable candidate to restore the $S_8$ ($\sigma_{8}$) tension.
The results of the joint analysis of \emph{Planck}+LSS are shown in the last two columns of Table \ref{tab:accumulate_constraints}.
When added with a prior on $S_8$ derived from joint LSS analysis, 
we find a $1\sigma$ detection of scattering between DM and photons and a larger upper limit of 
the cross section to mass ratio, i.e., $u_{\chi-\gamma} = 1.20^{+0.62}_{-0.80} \times 10^{-4}$ at $68\%$ C.L. and  $u_{\chi-\gamma} < 2.39\times
10^{-4} $ at $95\%$ C.L.
Meanwhile, the best-fit $S_8= 0.785\pm 0.017$ from the joint analysis closely matches the given $S_8$ prior, which suggests that
the cross section to mass ratio $u_{\chi-\gamma}$ are highly degenerate with $S_8$.
The addition of BOSS DR12 BAO and RSD ($f\sigma_{8}$) data have little impact on the fitting results. 
In conclusion, the interaction between DM and photons helps to reduce $S_8$ and thus is a possible solution to the $S_8$ tension.

\begin{figure*}
	\includegraphics[scale = 0.38]{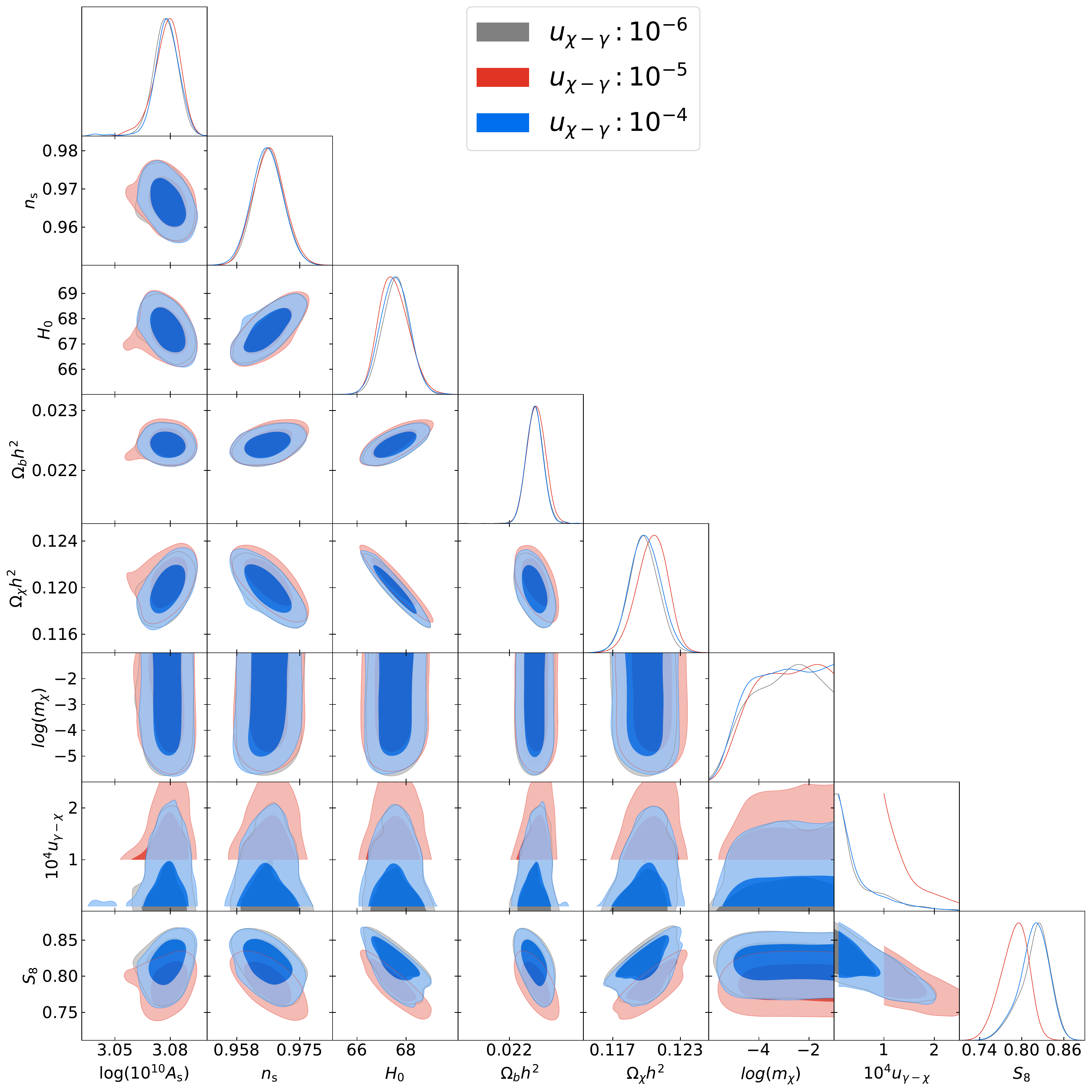}
	\caption{Constraints on cosmological parameters in the GDM scenario with different priors on $u_{\chi-\gamma}$ from 
		the \emph{Planck} 2018 low-$\ell$ TT+EE and high-$\ell$ TT+TE+EE power spectrum; 
		 The contours show $1\sigma$ and $2\sigma$ posterior distributions.
	}\label{fig:gdm_prior}
\end{figure*}

\subsection{Prior dependence}
As is discussed in previous sections, the parameter $m_\chi$ can be constrained by CMB through its impact on 
the EoS of DM. 
However, the EoS of DM
is also dependent on $u_{\chi-\gamma}$ which sets the decoupling scale ($a_{\rm{dec}}$) of DM-$\gamma$ interaction.
An obvious concern is whether the constrain on $m_\chi$ from CMB is dependent on the choice of $u_{\chi-\gamma}$ prior.
To account for that, we perform the fit to CMB datasets with different prior imposed on $u_{\chi-\gamma}$, i.e., the lower
limit of the uniform prior ranges from $10^{-4}$ to $10^{-6}$.
The posterior distributions are shown in Fig. \ref{fig:gdm_prior} and the parameter constraints are tabulated in Table \ref{tab:prior_constraints}.
It's obvious to see that, the different choice of  $u_{\chi-\gamma}$ priors have negligible impact on the constrains 
on $m_\chi$ as well as other cosmological parameters, except for the best-fit value of $S_8$.
As can be seen in the last column of Table \ref{tab:prior_constraints}, owing to the effect of diffusive damping the inferred value of $S_8$ is significantly reduced if we set the lower limit of $u_{\chi-\gamma}$ to be $10^{-4}$.
In summary, the mass of DM can be constrained independently of $u_{\chi-\gamma}$.

\section{CONCLUSIONS}\label{sec:conclusions}
The effect of DM-$\gamma$ interaction are usually parameterized by the dimensionless quantity $u_{\chi-\gamma}$ in former researches\cite{Wilkinson:2013kia,Stadler:2018jin}.
In this work, we investigate this scenario with an extra (new) parameter $m_\chi$, i.e., the masses of DM particles, 
which allows us to study the temperature evolution of DM ($T_\chi$) as well as its EoS ($w_\chi$). 
This scenario are often refferred to as generalized dark matter (GDM).
We have also studied the distinctive imprints of GDM on the matter power spectra and 
CMB temperature power spectra at both background and perturbation levels. These
distinctive imprints allow us to set a lower limit on $m_\chi$ within this scenario.
Due to the different cooling rates of DM and photons, the background bulk viscous pressure arises during the epoch of 
DM-$\gamma$ decoupling. However, the bulk viscous should only be counted DM and photons are treated as a single fluid. 
Our analysis suggests that, the peak value of bulk viscous pressure $\Pi$ is of the order of $10^{-1}P_\chi$ and 
thus have negligible impact on the background evolution of the universe.

We have modified the Boltzmann code \href{http://class-code.net/}{\texttt{CLASS}} to include the EoS of interacting DM, and use the \href{https://cobaya.readthedocs.io}{\texttt{Cobaya}} packages to perform the MCMC analysis.
In the fit of \emph{Planck} 2018 data, our numerical result yield comparable upper limit on $u_{\chi-\gamma}$ with previous 
works, i.e.,  $u_{\chi-\gamma} < 1.58 \times 10^{-4}$.
Also, one can effectively use \emph{Planck} data to restrict the mass of DM particles.
The restriction is robust against different priors on $u_{\chi-\gamma}$.
By fitting with the \emph{Planck} 2018 data alone we find $m_{\chi} > 8.7$keV at $95\%$ CL, while the inclusion of other
observational datasets does not significantly shift this result.
To test whether the GDM scenario restores the cosmic concordance, we perform the joint analysis of SH0ES+\emph{Planck} datasets.
The result suggests that $H_0$ only shifts very slightly and tensions with SH0ES remain at $2.5\sigma$.
This is understandable since the pressure of GDM has an effective presence only at very small scales and thus has negligible impact 
on $r_s(z_*)$ and the expansion history after last scattering.
Consequently, the GDM scenario is not likely to relieve the $H_0$ tension nor do the bulk viscous pressure which arises during the epoch of 
DM-$\gamma$ decoupling. 
The most prominent feature in the GDM scenario is the suppression in small-scale modes due to both the reduction in free-streaming monopole and the diffusive damping. 
When performing the joint analysis of \emph{Planck}+LSS datasets, the best-fit $S_8= 0.785\pm 0.017$ closely matches the given $S_8$ prior.
In summary, the GDM scenario should be counted as a viable candidate to restore the $S_8$ ($\sigma_{8}$) tension.\\

\appendix
\section{BULK VISCOUS}\label{sec:viscous_append}
We follow the key steps of the derivation in Ref. \cite{Zimdahl:1996fj}.
The coupled DM-$\gamma$ fluid is assumed to be at equilibrium at a certain time $\eta_0$, thus
\begin{subequations}
	\begin{align}
	T(\eta_0) &=  T_\gamma(\eta_0)= T_\chi(\eta_0), \\
	P(\eta_0) &= P_\gamma(\eta_0)+P_\chi(\eta_0),
	\end{align}\label{eq:equalibrum}
\end{subequations}
with $P_\gamma=n_\gamma k_B T_\gamma$, $ \rho_r=3 n_\gamma k_B T_r$ the pressure and energy density of photons, respectively.
During a subsequent time interval $\tau_c$, each component follows its own internal perfect fluid dynamics. Due to the different cooling rates of the components, there occur temperature differences between DM and photons $\Delta T \equiv T_\gamma-T_\chi$, which is expressed
as:
\begin{subequations}
	\begin{align}
	&T_\gamma-T_\chi=-3H \tau_c  T \left(\frac{\partial P_\gamma/\partial T}{\partial \rho_\gamma/\partial T }-\frac{\partial P_\chi/\partial T}{\partial \rho_\chi/\partial T }\right),\label{eq:T_diff} \\
	&T_\gamma-T=- 3H\tau_c  T \left(\frac{\partial P_\gamma/\partial T}{\partial \rho_\gamma/\partial T }-\frac{\partial P/\partial T}{\partial \rho/\partial T }\right), \\
	&T_\chi-T=-3H \tau_c  T \left(\frac{\partial P_\chi/\partial T}{\partial \rho_\chi/\partial T }-\frac{\partial P/\partial T}{\partial \rho/\partial T }\right).
	\end{align}
\end{subequations}
The temperature difference terms give rise to a bulk viscous pressure $\Pi$, i.e., 
\begin{equation}\label{totalpressure}
P_\gamma(n_\gamma,T_\gamma)+P_\chi(n_\chi,T_\chi)=P(n,T)+\Pi.
\end{equation} 
By definition $\Pi=-3H \xi$ with the bulk viscous coefficient $\xi$ is given by
\begin{eqnarray}
\xi &=& - \tau_c T \frac{\partial \rho}{\partial T}\left(\frac{\partial P_\gamma}{\partial \rho_\gamma}-\frac{\partial P}{\partial \rho}\right)\left(\frac{\partial P_\chi}{\partial \rho_\chi}-\frac{\partial P}{\partial \rho}\right), \\
&=& \tau_c \frac{n_\gamma k_B T}{3} \frac{n_{\chi}}{2 n_\gamma + n_{\chi}}. \label{eq:xi}  
\label{xi}
\end{eqnarray}
Noted that $\tau_c$ is defined as the time interval between the current collision and the next DM-$\gamma$ collision, after the last collision (when the two
fluids decouples from each other) $\tau_c$ should approach to zero. Conversely, when the two fluid are tightly coupled, $\tau_c$ should also approach to zero. According to
Eq. (\ref{eq:T_diff}), we can parameterize $\tau_c$ as:
\begin{eqnarray}
\tau_c = \frac{T_\gamma-T_\chi}{T H}\sigma(x),
\end{eqnarray} 
with $x\equiv\log(a/a_{\rm{dec}})$ and $\sigma(x) \equiv 1/(1+e^{-x})$ the sigmoid function. After DM-$\gamma$ decoupling ($a>a_{\rm{dec}}$) one has $\tau_c \sim 0$.
Here we have replaced $T$ with $T_\gamma$ according to the assumption of radiation dominance. Inserting
$\tau_c$ into Eq. (\ref{eq:xi}), the emergent bulk viscous can be expressed as:
\begin{eqnarray}
\Pi = -3H\xi = -\frac{1}{2+\delta} n_{\chi} k_B(T_\gamma-T_\chi)\sigma(x),
\end{eqnarray}  
where $\delta\equiv n_\chi/n_\gamma \approx 2.9/m_\chi[\rm{eV}/c^2]$.   
According to Eq. (\ref{eq:T_diff}), we can obtain the following relationship:
\begin{small}
	\begin{eqnarray}
	(T_\gamma-T)\frac{\partial P_\gamma}{\partial T} &=& \frac{1}{2+\delta} n_{\chi} k_B(T_\gamma-T_\chi)\sigma(x) = -\Pi,  \\
	(T_\chi-T)\frac{\partial P_\chi}{\partial T} &=&-\frac{2}{2+\delta} n_{\chi} k_B(T_\gamma-T_\chi)\sigma(x) =  2\Pi.
	\end{eqnarray}
\end{small}  
Accordingly, the sum of the fluid pressures can be expressed as
\begin{small}
	\begin{subequations}
		\begin{align}
		&P_\gamma(T_\gamma) + P_\chi(T_\chi)\label{eq:sum_pressure},\\ 
		=&P_\gamma(T) + P_\chi(T) + (T_\gamma-T)\frac{\partial P_\gamma}{\partial T}+(T_\chi-T)\frac{\partial P_\chi}{\partial T},\\
		=&P_\gamma(T) + P_\chi(T) -\Pi +2\Pi = p\:(T) +\Pi, \\ 
		=&P_\gamma(T_\gamma) + P_\chi(T) +2\Pi.\label{eq:sum_pressure2}
		\end{align}
	\end{subequations}
\end{small}
The bulk viscous arises as the DM temperature $T_\chi$ drops below the equilibrium temperature $T\approx T_\gamma$.
Comparing Eq. (\ref{eq:sum_pressure}) and Eq. (\ref{eq:sum_pressure2}) we obtain
\begin{eqnarray}
P_\chi(T_\chi) = P_\chi(T) +2\Pi.
\end{eqnarray}
This indicates that the (negative) bulk viscous pressure $\Pi$ can be viewed as a correction to 
overestimated DM pressure $P_\chi(T)$ when the two fluids are approximated by a single coupled fluid during a short period ($\tau_c$).

\begin{acknowledgments}
This work is supported in part by National Natural Science Foundation of China under Grant No. 12075042, Grant No. 11675032 (People's Republic of China).
\end{acknowledgments}


\bibliography{gdm}

\begin{thebibliography}{82}%
\makeatletter
\providecommand \@ifxundefined [1]{%
 \@ifx{#1\undefined}
}%
\providecommand \@ifnum [1]{%
 \ifnum #1\expandafter \@firstoftwo
 \else \expandafter \@secondoftwo
 \fi
}%
\providecommand \@ifx [1]{%
 \ifx #1\expandafter \@firstoftwo
 \else \expandafter \@secondoftwo
 \fi
}%
\providecommand \natexlab [1]{#1}%
\providecommand \enquote  [1]{``#1''}%
\providecommand \bibnamefont  [1]{#1}%
\providecommand \bibfnamefont [1]{#1}%
\providecommand \citenamefont [1]{#1}%
\providecommand \href@noop [0]{\@secondoftwo}%
\providecommand \href [0]{\begingroup \@sanitize@url \@href}%
\providecommand \@href[1]{\@@startlink{#1}\@@href}%
\providecommand \@@href[1]{\endgroup#1\@@endlink}%
\providecommand \@sanitize@url [0]{\catcode `\\12\catcode `\$12\catcode
  `\&12\catcode `\#12\catcode `\^12\catcode `\_12\catcode `\%12\relax}%
\providecommand \@@startlink[1]{}%
\providecommand \@@endlink[0]{}%
\providecommand \url  [0]{\begingroup\@sanitize@url \@url }%
\providecommand \@url [1]{\endgroup\@href {#1}{\urlprefix }}%
\providecommand \urlprefix  [0]{URL }%
\providecommand \Eprint [0]{\href }%
\providecommand \doibase [0]{http://dx.doi.org/}%
\providecommand \selectlanguage [0]{\@gobble}%
\providecommand \bibinfo  [0]{\@secondoftwo}%
\providecommand \bibfield  [0]{\@secondoftwo}%
\providecommand \translation [1]{[#1]}%
\providecommand \BibitemOpen [0]{}%
\providecommand \bibitemStop [0]{}%
\providecommand \bibitemNoStop [0]{.\EOS\space}%
\providecommand \EOS [0]{\spacefactor3000\relax}%
\providecommand \BibitemShut  [1]{\csname bibitem#1\endcsname}%
\let\auto@bib@innerbib\@empty
\bibitem [{\citenamefont {Riess}\ \emph {et~al.}(1998)\citenamefont {Riess}
  \emph {et~al.}}]{SupernovaSearchTeam:1998fmf}%
  \BibitemOpen
  \bibfield  {author} {\bibinfo {author} {\bibfnamefont {A.~G.}\ \bibnamefont
  {Riess}} \emph {et~al.} (\bibinfo {collaboration} {Supernova Search Team}),\
  }\href {\doibase 10.1086/300499} {\bibfield  {journal} {\bibinfo  {journal}
  {Astron. J.}\ }\textbf {\bibinfo {volume} {116}},\ \bibinfo {pages} {1009}
  (\bibinfo {year} {1998})},\ \Eprint {http://arxiv.org/abs/astro-ph/9805201}
  {arXiv:astro-ph/9805201} \BibitemShut {NoStop}%
\bibitem [{\citenamefont {Aghanim}\ \emph {et~al.}(2018)\citenamefont {Aghanim}
  \emph {et~al.}}]{Aghanim:2018eyx}%
  \BibitemOpen
  \bibfield  {author} {\bibinfo {author} {\bibfnamefont {N.}~\bibnamefont
  {Aghanim}} \emph {et~al.} (\bibinfo {collaboration} {Planck}),\ }\href@noop
  {} {\  (\bibinfo {year} {2018})},\ \Eprint {http://arxiv.org/abs/1807.06209}
  {arXiv:1807.06209 [astro-ph.CO]} \BibitemShut {NoStop}%
\bibitem [{\citenamefont {Aghanim}\ \emph
  {et~al.}(2020{\natexlab{a}})\citenamefont {Aghanim} \emph
  {et~al.}}]{Planck:2019nip}%
  \BibitemOpen
  \bibfield  {author} {\bibinfo {author} {\bibfnamefont {N.}~\bibnamefont
  {Aghanim}} \emph {et~al.} (\bibinfo {collaboration} {Planck}),\ }\href
  {\doibase 10.1051/0004-6361/201936386} {\bibfield  {journal} {\bibinfo
  {journal} {Astron. Astrophys.}\ }\textbf {\bibinfo {volume} {641}},\ \bibinfo
  {pages} {A5} (\bibinfo {year} {2020}{\natexlab{a}})},\ \Eprint
  {http://arxiv.org/abs/1907.12875} {arXiv:1907.12875 [astro-ph.CO]}
  \BibitemShut {NoStop}%
\bibitem [{\citenamefont {Abbott}\ \emph {et~al.}(2018)\citenamefont {Abbott}
  \emph {et~al.}}]{Abbott:2017wau}%
  \BibitemOpen
  \bibfield  {author} {\bibinfo {author} {\bibfnamefont {T.}~\bibnamefont
  {Abbott}} \emph {et~al.} (\bibinfo {collaboration} {DES}),\ }\href {\doibase
  10.1103/PhysRevD.98.043526} {\bibfield  {journal} {\bibinfo  {journal} {Phys.
  Rev. D}\ }\textbf {\bibinfo {volume} {98}},\ \bibinfo {pages} {043526}
  (\bibinfo {year} {2018})},\ \Eprint {http://arxiv.org/abs/1708.01530}
  {arXiv:1708.01530 [astro-ph.CO]} \BibitemShut {NoStop}%
\bibitem [{\citenamefont {Alam}\ \emph {et~al.}(2017)\citenamefont {Alam} \emph
  {et~al.}}]{Alam:2016hwk}%
  \BibitemOpen
  \bibfield  {author} {\bibinfo {author} {\bibfnamefont {S.}~\bibnamefont
  {Alam}} \emph {et~al.} (\bibinfo {collaboration} {BOSS}),\ }\href {\doibase
  10.1093/mnras/stx721} {\bibfield  {journal} {\bibinfo  {journal} {Mon. Not.
  Roy. Astron. Soc.}\ }\textbf {\bibinfo {volume} {470}},\ \bibinfo {pages}
  {2617} (\bibinfo {year} {2017})},\ \Eprint {http://arxiv.org/abs/1607.03155}
  {arXiv:1607.03155 [astro-ph.CO]} \BibitemShut {NoStop}%
\bibitem [{\citenamefont {Li}\ \emph {et~al.}(2011)\citenamefont {Li},
  \citenamefont {Li}, \citenamefont {Wang},\ and\ \citenamefont
  {Wang}}]{Li_2011}%
  \BibitemOpen
  \bibfield  {author} {\bibinfo {author} {\bibfnamefont {M.}~\bibnamefont
  {Li}}, \bibinfo {author} {\bibfnamefont {X.-D.}\ \bibnamefont {Li}}, \bibinfo
  {author} {\bibfnamefont {S.}~\bibnamefont {Wang}}, \ and\ \bibinfo {author}
  {\bibfnamefont {Y.}~\bibnamefont {Wang}},\ }\href {\doibase
  10.1088/0253-6102/56/3/24} {\bibfield  {journal} {\bibinfo  {journal}
  {Communications in Theoretical Physics}\ }\textbf {\bibinfo {volume} {56}},\
  \bibinfo {pages} {525} (\bibinfo {year} {2011})}\BibitemShut {NoStop}%
\bibitem [{\citenamefont {Persic}\ \emph {et~al.}(1996)\citenamefont {Persic},
  \citenamefont {Salucci},\ and\ \citenamefont {Stel}}]{Persic:1995ru}%
  \BibitemOpen
  \bibfield  {author} {\bibinfo {author} {\bibfnamefont {M.}~\bibnamefont
  {Persic}}, \bibinfo {author} {\bibfnamefont {P.}~\bibnamefont {Salucci}}, \
  and\ \bibinfo {author} {\bibfnamefont {F.}~\bibnamefont {Stel}},\ }\href
  {\doibase 10.1093/mnras/278.1.27} {\bibfield  {journal} {\bibinfo  {journal}
  {Mon. Not. Roy. Astron. Soc.}\ }\textbf {\bibinfo {volume} {281}},\ \bibinfo
  {pages} {27} (\bibinfo {year} {1996})},\ \Eprint
  {http://arxiv.org/abs/astro-ph/9506004} {arXiv:astro-ph/9506004} \BibitemShut
  {NoStop}%
\bibitem [{\citenamefont {Clowe}\ \emph {et~al.}(2006)\citenamefont {Clowe},
  \citenamefont {Bradac}, \citenamefont {Gonzalez}, \citenamefont {Markevitch},
  \citenamefont {Randall}, \citenamefont {Jones},\ and\ \citenamefont
  {Zaritsky}}]{Clowe:2006eq}%
  \BibitemOpen
  \bibfield  {author} {\bibinfo {author} {\bibfnamefont {D.}~\bibnamefont
  {Clowe}}, \bibinfo {author} {\bibfnamefont {M.}~\bibnamefont {Bradac}},
  \bibinfo {author} {\bibfnamefont {A.~H.}\ \bibnamefont {Gonzalez}}, \bibinfo
  {author} {\bibfnamefont {M.}~\bibnamefont {Markevitch}}, \bibinfo {author}
  {\bibfnamefont {S.~W.}\ \bibnamefont {Randall}}, \bibinfo {author}
  {\bibfnamefont {C.}~\bibnamefont {Jones}}, \ and\ \bibinfo {author}
  {\bibfnamefont {D.}~\bibnamefont {Zaritsky}},\ }\href {\doibase
  10.1086/508162} {\bibfield  {journal} {\bibinfo  {journal} {Astrophys. J.
  Lett.}\ }\textbf {\bibinfo {volume} {648}},\ \bibinfo {pages} {L109}
  (\bibinfo {year} {2006})},\ \Eprint {http://arxiv.org/abs/astro-ph/0608407}
  {arXiv:astro-ph/0608407} \BibitemShut {NoStop}%
\bibitem [{\citenamefont {Harvey}\ \emph {et~al.}(2015)\citenamefont {Harvey},
  \citenamefont {Massey}, \citenamefont {Kitching}, \citenamefont {Taylor},\
  and\ \citenamefont {Tittley}}]{Harvey:2015hha}%
  \BibitemOpen
  \bibfield  {author} {\bibinfo {author} {\bibfnamefont {D.}~\bibnamefont
  {Harvey}}, \bibinfo {author} {\bibfnamefont {R.}~\bibnamefont {Massey}},
  \bibinfo {author} {\bibfnamefont {T.}~\bibnamefont {Kitching}}, \bibinfo
  {author} {\bibfnamefont {A.}~\bibnamefont {Taylor}}, \ and\ \bibinfo {author}
  {\bibfnamefont {E.}~\bibnamefont {Tittley}},\ }\href {\doibase
  10.1126/science.1261381} {\bibfield  {journal} {\bibinfo  {journal}
  {Science}\ }\textbf {\bibinfo {volume} {347}},\ \bibinfo {pages} {1462}
  (\bibinfo {year} {2015})},\ \Eprint {http://arxiv.org/abs/1503.07675}
  {arXiv:1503.07675 [astro-ph.CO]} \BibitemShut {NoStop}%
\bibitem [{\citenamefont {Griest}\ and\ \citenamefont
  {Kamionkowski}(1990)}]{griest1990unitarity}%
  \BibitemOpen
  \bibfield  {author} {\bibinfo {author} {\bibfnamefont {K.}~\bibnamefont
  {Griest}}\ and\ \bibinfo {author} {\bibfnamefont {M.}~\bibnamefont
  {Kamionkowski}},\ }\href@noop {} {\bibfield  {journal} {\bibinfo  {journal}
  {Physical Review Letters}\ }\textbf {\bibinfo {volume} {64}},\ \bibinfo
  {pages} {615} (\bibinfo {year} {1990})}\BibitemShut {NoStop}%
\bibitem [{\citenamefont {Boehm}\ \emph {et~al.}(2001)\citenamefont {Boehm},
  \citenamefont {Fayet},\ and\ \citenamefont {Schaeffer}}]{Boehm:2000gq}%
  \BibitemOpen
  \bibfield  {author} {\bibinfo {author} {\bibfnamefont {C.}~\bibnamefont
  {Boehm}}, \bibinfo {author} {\bibfnamefont {P.}~\bibnamefont {Fayet}}, \ and\
  \bibinfo {author} {\bibfnamefont {R.}~\bibnamefont {Schaeffer}},\ }\href
  {\doibase 10.1016/S0370-2693(01)01060-7} {\bibfield  {journal} {\bibinfo
  {journal} {Phys. Lett. B}\ }\textbf {\bibinfo {volume} {518}},\ \bibinfo
  {pages} {8} (\bibinfo {year} {2001})},\ \Eprint
  {http://arxiv.org/abs/astro-ph/0012504} {arXiv:astro-ph/0012504} \BibitemShut
  {NoStop}%
\bibitem [{\citenamefont {Boyarsky}\ \emph {et~al.}(2009)\citenamefont
  {Boyarsky}, \citenamefont {Ruchayskiy},\ and\ \citenamefont
  {Iakubovskyi}}]{Boyarsky:2008ju}%
  \BibitemOpen
  \bibfield  {author} {\bibinfo {author} {\bibfnamefont {A.}~\bibnamefont
  {Boyarsky}}, \bibinfo {author} {\bibfnamefont {O.}~\bibnamefont
  {Ruchayskiy}}, \ and\ \bibinfo {author} {\bibfnamefont {D.}~\bibnamefont
  {Iakubovskyi}},\ }\href {\doibase 10.1088/1475-7516/2009/03/005} {\bibfield
  {journal} {\bibinfo  {journal} {JCAP}\ }\textbf {\bibinfo {volume} {03}},\
  \bibinfo {pages} {005} (\bibinfo {year} {2009})},\ \Eprint
  {http://arxiv.org/abs/0808.3902} {arXiv:0808.3902 [hep-ph]} \BibitemShut
  {NoStop}%
\bibitem [{\citenamefont {Kavanagh}\ and\ \citenamefont
  {Green}(2013)}]{Kavanagh:2013wba}%
  \BibitemOpen
  \bibfield  {author} {\bibinfo {author} {\bibfnamefont {B.~J.}\ \bibnamefont
  {Kavanagh}}\ and\ \bibinfo {author} {\bibfnamefont {A.~M.}\ \bibnamefont
  {Green}},\ }\href {\doibase 10.1103/PhysRevLett.111.031302} {\bibfield
  {journal} {\bibinfo  {journal} {Phys. Rev. Lett.}\ }\textbf {\bibinfo
  {volume} {111}},\ \bibinfo {pages} {031302} (\bibinfo {year} {2013})},\
  \Eprint {http://arxiv.org/abs/1303.6868} {arXiv:1303.6868 [astro-ph.CO]}
  \BibitemShut {NoStop}%
\bibitem [{\citenamefont {Berlin}\ and\ \citenamefont
  {Blinov}(2018)}]{Berlin:2017ftj}%
  \BibitemOpen
  \bibfield  {author} {\bibinfo {author} {\bibfnamefont {A.}~\bibnamefont
  {Berlin}}\ and\ \bibinfo {author} {\bibfnamefont {N.}~\bibnamefont
  {Blinov}},\ }\href {\doibase 10.1103/PhysRevLett.120.021801} {\bibfield
  {journal} {\bibinfo  {journal} {Phys. Rev. Lett.}\ }\textbf {\bibinfo
  {volume} {120}},\ \bibinfo {pages} {021801} (\bibinfo {year} {2018})},\
  \Eprint {http://arxiv.org/abs/1706.07046} {arXiv:1706.07046 [hep-ph]}
  \BibitemShut {NoStop}%
\bibitem [{\citenamefont {Agnes}\ \emph {et~al.}(2018)\citenamefont {Agnes}
  \emph {et~al.}}]{DarkSide:2018bpj}%
  \BibitemOpen
  \bibfield  {author} {\bibinfo {author} {\bibfnamefont {P.}~\bibnamefont
  {Agnes}} \emph {et~al.} (\bibinfo {collaboration} {DarkSide}),\ }\href
  {\doibase 10.1103/PhysRevLett.121.081307} {\bibfield  {journal} {\bibinfo
  {journal} {Phys. Rev. Lett.}\ }\textbf {\bibinfo {volume} {121}},\ \bibinfo
  {pages} {081307} (\bibinfo {year} {2018})},\ \Eprint
  {http://arxiv.org/abs/1802.06994} {arXiv:1802.06994 [astro-ph.HE]}
  \BibitemShut {NoStop}%
\bibitem [{\citenamefont {Perivolaropoulos}\ and\ \citenamefont
  {Skara}(2021)}]{Perivolaropoulos:2021jda}%
  \BibitemOpen
  \bibfield  {author} {\bibinfo {author} {\bibfnamefont {L.}~\bibnamefont
  {Perivolaropoulos}}\ and\ \bibinfo {author} {\bibfnamefont {F.}~\bibnamefont
  {Skara}},\ }\href@noop {} {\  (\bibinfo {year} {2021})},\ \Eprint
  {http://arxiv.org/abs/2105.05208} {arXiv:2105.05208 [astro-ph.CO]}
  \BibitemShut {NoStop}%
\bibitem [{\citenamefont {Verde}\ \emph {et~al.}(2019)\citenamefont {Verde},
  \citenamefont {Treu},\ and\ \citenamefont {Riess}}]{Verde:2019ivm}%
  \BibitemOpen
  \bibfield  {author} {\bibinfo {author} {\bibfnamefont {L.}~\bibnamefont
  {Verde}}, \bibinfo {author} {\bibfnamefont {T.}~\bibnamefont {Treu}}, \ and\
  \bibinfo {author} {\bibfnamefont {A.}~\bibnamefont {Riess}},\ }\href
  {\doibase 10.1038/s41550-019-0902-0} {\bibfield  {journal} {\bibinfo
  {journal} {Nature Astron.}\ }\textbf {\bibinfo {volume} {3}},\ \bibinfo
  {pages} {891} (\bibinfo {year} {2019})},\ \Eprint
  {http://arxiv.org/abs/1907.10625} {arXiv:1907.10625 [astro-ph.CO]}
  \BibitemShut {NoStop}%
\bibitem [{\citenamefont {Di~Valentino}\ \emph {et~al.}(2021)\citenamefont
  {Di~Valentino}, \citenamefont {Mena}, \citenamefont {Pan}, \citenamefont
  {Visinelli}, \citenamefont {Yang}, \citenamefont {Melchiorri}, \citenamefont
  {Mota}, \citenamefont {Riess},\ and\ \citenamefont
  {Silk}}]{DiValentino:2021izs}%
  \BibitemOpen
  \bibfield  {author} {\bibinfo {author} {\bibfnamefont {E.}~\bibnamefont
  {Di~Valentino}}, \bibinfo {author} {\bibfnamefont {O.}~\bibnamefont {Mena}},
  \bibinfo {author} {\bibfnamefont {S.}~\bibnamefont {Pan}}, \bibinfo {author}
  {\bibfnamefont {L.}~\bibnamefont {Visinelli}}, \bibinfo {author}
  {\bibfnamefont {W.}~\bibnamefont {Yang}}, \bibinfo {author} {\bibfnamefont
  {A.}~\bibnamefont {Melchiorri}}, \bibinfo {author} {\bibfnamefont {D.~F.}\
  \bibnamefont {Mota}}, \bibinfo {author} {\bibfnamefont {A.~G.}\ \bibnamefont
  {Riess}}, \ and\ \bibinfo {author} {\bibfnamefont {J.}~\bibnamefont {Silk}},\
  }\href {\doibase 10.1088/1361-6382/ac086d} {\bibfield  {journal} {\bibinfo
  {journal} {Class. Quant. Grav.}\ }\textbf {\bibinfo {volume} {38}},\ \bibinfo
  {pages} {153001} (\bibinfo {year} {2021})},\ \Eprint
  {http://arxiv.org/abs/2103.01183} {arXiv:2103.01183 [astro-ph.CO]}
  \BibitemShut {NoStop}%
\bibitem [{\citenamefont {Hildebrandt}\ \emph {et~al.}(2017)\citenamefont
  {Hildebrandt} \emph {et~al.}}]{Hildebrandt:2016iqg}%
  \BibitemOpen
  \bibfield  {author} {\bibinfo {author} {\bibfnamefont {H.}~\bibnamefont
  {Hildebrandt}} \emph {et~al.},\ }\href {\doibase 10.1093/mnras/stw2805}
  {\bibfield  {journal} {\bibinfo  {journal} {Mon. Not. Roy. Astron. Soc.}\
  }\textbf {\bibinfo {volume} {465}},\ \bibinfo {pages} {1454} (\bibinfo {year}
  {2017})},\ \Eprint {http://arxiv.org/abs/1606.05338} {arXiv:1606.05338
  [astro-ph.CO]} \BibitemShut {NoStop}%
\bibitem [{\citenamefont {Hildebrandt}\ \emph {et~al.}(2020)\citenamefont
  {Hildebrandt} \emph {et~al.}}]{Hildebrandt:2018yau}%
  \BibitemOpen
  \bibfield  {author} {\bibinfo {author} {\bibfnamefont {H.}~\bibnamefont
  {Hildebrandt}} \emph {et~al.},\ }\href {\doibase 10.1051/0004-6361/201834878}
  {\bibfield  {journal} {\bibinfo  {journal} {Astron. Astrophys.}\ }\textbf
  {\bibinfo {volume} {633}},\ \bibinfo {pages} {A69} (\bibinfo {year}
  {2020})},\ \Eprint {http://arxiv.org/abs/1812.06076} {arXiv:1812.06076
  [astro-ph.CO]} \BibitemShut {NoStop}%
\bibitem [{\citenamefont {Hikage}\ \emph {et~al.}(2019)\citenamefont {Hikage}
  \emph {et~al.}}]{HSC:2018mrq}%
  \BibitemOpen
  \bibfield  {author} {\bibinfo {author} {\bibfnamefont {C.}~\bibnamefont
  {Hikage}} \emph {et~al.} (\bibinfo {collaboration} {HSC}),\ }\href {\doibase
  10.1093/pasj/psz010} {\bibfield  {journal} {\bibinfo  {journal} {Publ.
  Astron. Soc. Jap.}\ }\textbf {\bibinfo {volume} {71}},\ \bibinfo {pages} {43}
  (\bibinfo {year} {2019})},\ \Eprint {http://arxiv.org/abs/1809.09148}
  {arXiv:1809.09148 [astro-ph.CO]} \BibitemShut {NoStop}%
\bibitem [{\citenamefont {Bellazzini}\ \emph {et~al.}(2013)\citenamefont
  {Bellazzini}, \citenamefont {Cliche},\ and\ \citenamefont
  {Tanedo}}]{Bellazzini:2013foa}%
  \BibitemOpen
  \bibfield  {author} {\bibinfo {author} {\bibfnamefont {B.}~\bibnamefont
  {Bellazzini}}, \bibinfo {author} {\bibfnamefont {M.}~\bibnamefont {Cliche}},
  \ and\ \bibinfo {author} {\bibfnamefont {P.}~\bibnamefont {Tanedo}},\ }\href
  {\doibase 10.1103/PhysRevD.88.083506} {\bibfield  {journal} {\bibinfo
  {journal} {Phys. Rev. D}\ }\textbf {\bibinfo {volume} {88}},\ \bibinfo
  {pages} {083506} (\bibinfo {year} {2013})},\ \Eprint
  {http://arxiv.org/abs/1307.1129} {arXiv:1307.1129 [hep-ph]} \BibitemShut
  {NoStop}%
\bibitem [{\citenamefont {Tulin}\ and\ \citenamefont {Yu}(2018)}]{TULIN20181}%
  \BibitemOpen
  \bibfield  {author} {\bibinfo {author} {\bibfnamefont {S.}~\bibnamefont
  {Tulin}}\ and\ \bibinfo {author} {\bibfnamefont {H.-B.}\ \bibnamefont {Yu}},\
  }\href {\doibase https://doi.org/10.1016/j.physrep.2017.11.004} {\bibfield
  {journal} {\bibinfo  {journal} {Physics Reports}\ }\textbf {\bibinfo {volume}
  {730}},\ \bibinfo {pages} {1} (\bibinfo {year} {2018})},\ \bibinfo {note}
  {dark matter self-interactions and small scale structure}\BibitemShut
  {NoStop}%
\bibitem [{\citenamefont {Salucci}(2019)}]{Salucci:2018hqu}%
  \BibitemOpen
  \bibfield  {author} {\bibinfo {author} {\bibfnamefont {P.}~\bibnamefont
  {Salucci}},\ }\href {\doibase 10.1007/s00159-018-0113-1} {\bibfield
  {journal} {\bibinfo  {journal} {Astron. Astrophys. Rev.}\ }\textbf {\bibinfo
  {volume} {27}},\ \bibinfo {pages} {2} (\bibinfo {year} {2019})},\ \Eprint
  {http://arxiv.org/abs/1811.08843} {arXiv:1811.08843 [astro-ph.GA]}
  \BibitemShut {NoStop}%
\bibitem [{\citenamefont {Barkana}(2018)}]{Barkana:2018lgd}%
  \BibitemOpen
  \bibfield  {author} {\bibinfo {author} {\bibfnamefont {R.}~\bibnamefont
  {Barkana}},\ }\href {\doibase 10.1038/nature25791} {\bibfield  {journal}
  {\bibinfo  {journal} {Nature}\ }\textbf {\bibinfo {volume} {555}},\ \bibinfo
  {pages} {71} (\bibinfo {year} {2018})},\ \Eprint
  {http://arxiv.org/abs/1803.06698} {arXiv:1803.06698 [astro-ph.CO]}
  \BibitemShut {NoStop}%
\bibitem [{\citenamefont {Mu\~noz}\ and\ \citenamefont
  {Loeb}(2017)}]{Munoz:2017qpy}%
  \BibitemOpen
  \bibfield  {author} {\bibinfo {author} {\bibfnamefont {J.~B.}\ \bibnamefont
  {Mu\~noz}}\ and\ \bibinfo {author} {\bibfnamefont {A.}~\bibnamefont {Loeb}},\
  }\href {\doibase 10.1088/1475-7516/2017/11/043} {\bibfield  {journal}
  {\bibinfo  {journal} {JCAP}\ }\textbf {\bibinfo {volume} {11}},\ \bibinfo
  {pages} {043} (\bibinfo {year} {2017})},\ \Eprint
  {http://arxiv.org/abs/1708.08923} {arXiv:1708.08923 [astro-ph.CO]}
  \BibitemShut {NoStop}%
\bibitem [{\citenamefont {Ali-Haimoud}\ \emph {et~al.}(2015)\citenamefont
  {Ali-Haimoud}, \citenamefont {Chluba},\ and\ \citenamefont
  {Kamionkowski}}]{Ali-Haimoud:2015pwa}%
  \BibitemOpen
  \bibfield  {author} {\bibinfo {author} {\bibfnamefont {Y.}~\bibnamefont
  {Ali-Haimoud}}, \bibinfo {author} {\bibfnamefont {J.}~\bibnamefont {Chluba}},
  \ and\ \bibinfo {author} {\bibfnamefont {M.}~\bibnamefont {Kamionkowski}},\
  }\href {\doibase 10.1103/PhysRevLett.115.071304} {\bibfield  {journal}
  {\bibinfo  {journal} {Phys. Rev. Lett.}\ }\textbf {\bibinfo {volume} {115}},\
  \bibinfo {pages} {071304} (\bibinfo {year} {2015})},\ \Eprint
  {http://arxiv.org/abs/1506.04745} {arXiv:1506.04745 [astro-ph.CO]}
  \BibitemShut {NoStop}%
\bibitem [{\citenamefont {Mu\~noz}\ \emph {et~al.}(2015)\citenamefont
  {Mu\~noz}, \citenamefont {Kovetz},\ and\ \citenamefont
  {Ali-Ha\"\i{}moud}}]{Munoz:2015bca}%
  \BibitemOpen
  \bibfield  {author} {\bibinfo {author} {\bibfnamefont {J.~B.}\ \bibnamefont
  {Mu\~noz}}, \bibinfo {author} {\bibfnamefont {E.~D.}\ \bibnamefont {Kovetz}},
  \ and\ \bibinfo {author} {\bibfnamefont {Y.}~\bibnamefont
  {Ali-Ha\"\i{}moud}},\ }\href {\doibase 10.1103/PhysRevD.92.083528} {\bibfield
   {journal} {\bibinfo  {journal} {Phys. Rev. D}\ }\textbf {\bibinfo {volume}
  {92}},\ \bibinfo {pages} {083528} (\bibinfo {year} {2015})},\ \Eprint
  {http://arxiv.org/abs/1509.00029} {arXiv:1509.00029 [astro-ph.CO]}
  \BibitemShut {NoStop}%
\bibitem [{\citenamefont {Gluscevic}\ and\ \citenamefont
  {Boddy}(2018)}]{Gluscevic:2017ywp}%
  \BibitemOpen
  \bibfield  {author} {\bibinfo {author} {\bibfnamefont {V.}~\bibnamefont
  {Gluscevic}}\ and\ \bibinfo {author} {\bibfnamefont {K.~K.}\ \bibnamefont
  {Boddy}},\ }\href {\doibase 10.1103/PhysRevLett.121.081301} {\bibfield
  {journal} {\bibinfo  {journal} {Phys. Rev. Lett.}\ }\textbf {\bibinfo
  {volume} {121}},\ \bibinfo {pages} {081301} (\bibinfo {year} {2018})},\
  \Eprint {http://arxiv.org/abs/1712.07133} {arXiv:1712.07133 [astro-ph.CO]}
  \BibitemShut {NoStop}%
\bibitem [{\citenamefont {Slatyer}\ and\ \citenamefont
  {Wu}(2018)}]{Slatyer:2018aqg}%
  \BibitemOpen
  \bibfield  {author} {\bibinfo {author} {\bibfnamefont {T.~R.}\ \bibnamefont
  {Slatyer}}\ and\ \bibinfo {author} {\bibfnamefont {C.-L.}\ \bibnamefont
  {Wu}},\ }\href {\doibase 10.1103/PhysRevD.98.023013} {\bibfield  {journal}
  {\bibinfo  {journal} {Phys. Rev. D}\ }\textbf {\bibinfo {volume} {98}},\
  \bibinfo {pages} {023013} (\bibinfo {year} {2018})},\ \Eprint
  {http://arxiv.org/abs/1803.09734} {arXiv:1803.09734 [astro-ph.CO]}
  \BibitemShut {NoStop}%
\bibitem [{\citenamefont {Xu}\ \emph {et~al.}(2018)\citenamefont {Xu},
  \citenamefont {Dvorkin},\ and\ \citenamefont {Chael}}]{Xu:2018efh}%
  \BibitemOpen
  \bibfield  {author} {\bibinfo {author} {\bibfnamefont {W.~L.}\ \bibnamefont
  {Xu}}, \bibinfo {author} {\bibfnamefont {C.}~\bibnamefont {Dvorkin}}, \ and\
  \bibinfo {author} {\bibfnamefont {A.}~\bibnamefont {Chael}},\ }\href
  {\doibase 10.1103/PhysRevD.97.103530} {\bibfield  {journal} {\bibinfo
  {journal} {Phys. Rev. D}\ }\textbf {\bibinfo {volume} {97}},\ \bibinfo
  {pages} {103530} (\bibinfo {year} {2018})},\ \Eprint
  {http://arxiv.org/abs/1802.06788} {arXiv:1802.06788 [astro-ph.CO]}
  \BibitemShut {NoStop}%
\bibitem [{\citenamefont {Bringmann}\ \emph {et~al.}(2014)\citenamefont
  {Bringmann}, \citenamefont {Hasenkamp},\ and\ \citenamefont
  {Kersten}}]{Bringmann:2013vra}%
  \BibitemOpen
  \bibfield  {author} {\bibinfo {author} {\bibfnamefont {T.}~\bibnamefont
  {Bringmann}}, \bibinfo {author} {\bibfnamefont {J.}~\bibnamefont
  {Hasenkamp}}, \ and\ \bibinfo {author} {\bibfnamefont {J.}~\bibnamefont
  {Kersten}},\ }\href {\doibase 10.1088/1475-7516/2014/07/042} {\bibfield
  {journal} {\bibinfo  {journal} {JCAP}\ }\textbf {\bibinfo {volume} {07}},\
  \bibinfo {pages} {042} (\bibinfo {year} {2014})},\ \Eprint
  {http://arxiv.org/abs/1312.4947} {arXiv:1312.4947 [hep-ph]} \BibitemShut
  {NoStop}%
\bibitem [{\citenamefont {Audren}\ \emph {et~al.}(2015)\citenamefont {Audren}
  \emph {et~al.}}]{Audren:2014lsa}%
  \BibitemOpen
  \bibfield  {author} {\bibinfo {author} {\bibfnamefont {B.}~\bibnamefont
  {Audren}} \emph {et~al.},\ }\href {\doibase 10.1088/1475-7516/2015/03/036}
  {\bibfield  {journal} {\bibinfo  {journal} {JCAP}\ }\textbf {\bibinfo
  {volume} {03}},\ \bibinfo {pages} {036} (\bibinfo {year} {2015})},\ \Eprint
  {http://arxiv.org/abs/1412.5948} {arXiv:1412.5948 [astro-ph.CO]} \BibitemShut
  {NoStop}%
\bibitem [{\citenamefont {Horiuchi}\ \emph {et~al.}(2016)\citenamefont
  {Horiuchi}, \citenamefont {Bozek}, \citenamefont {Abazajian}, \citenamefont
  {Boylan-Kolchin}, \citenamefont {Bullock}, \citenamefont {Garrison-Kimmel},\
  and\ \citenamefont {Onorbe}}]{Horiuchi:2015qri}%
  \BibitemOpen
  \bibfield  {author} {\bibinfo {author} {\bibfnamefont {S.}~\bibnamefont
  {Horiuchi}}, \bibinfo {author} {\bibfnamefont {B.}~\bibnamefont {Bozek}},
  \bibinfo {author} {\bibfnamefont {K.~N.}\ \bibnamefont {Abazajian}}, \bibinfo
  {author} {\bibfnamefont {M.}~\bibnamefont {Boylan-Kolchin}}, \bibinfo
  {author} {\bibfnamefont {J.~S.}\ \bibnamefont {Bullock}}, \bibinfo {author}
  {\bibfnamefont {S.}~\bibnamefont {Garrison-Kimmel}}, \ and\ \bibinfo {author}
  {\bibfnamefont {J.}~\bibnamefont {Onorbe}},\ }\href {\doibase
  10.1093/mnras/stv2922} {\bibfield  {journal} {\bibinfo  {journal} {Mon. Not.
  Roy. Astron. Soc.}\ }\textbf {\bibinfo {volume} {456}},\ \bibinfo {pages}
  {4346} (\bibinfo {year} {2016})},\ \Eprint {http://arxiv.org/abs/1512.04548}
  {arXiv:1512.04548 [astro-ph.CO]} \BibitemShut {NoStop}%
\bibitem [{\citenamefont {Di~Valentino}\ \emph {et~al.}(2018)\citenamefont
  {Di~Valentino}, \citenamefont {B\o{}ehm}, \citenamefont {Hivon},\ and\
  \citenamefont {Bouchet}}]{DiValentino:2017oaw}%
  \BibitemOpen
  \bibfield  {author} {\bibinfo {author} {\bibfnamefont {E.}~\bibnamefont
  {Di~Valentino}}, \bibinfo {author} {\bibfnamefont {C.}~\bibnamefont
  {B\o{}ehm}}, \bibinfo {author} {\bibfnamefont {E.}~\bibnamefont {Hivon}}, \
  and\ \bibinfo {author} {\bibfnamefont {F.~R.}\ \bibnamefont {Bouchet}},\
  }\href {\doibase 10.1103/PhysRevD.97.043513} {\bibfield  {journal} {\bibinfo
  {journal} {Phys. Rev. D}\ }\textbf {\bibinfo {volume} {97}},\ \bibinfo
  {pages} {043513} (\bibinfo {year} {2018})},\ \Eprint
  {http://arxiv.org/abs/1710.02559} {arXiv:1710.02559 [astro-ph.CO]}
  \BibitemShut {NoStop}%
\bibitem [{\citenamefont {Pandey}\ \emph {et~al.}(2019)\citenamefont {Pandey},
  \citenamefont {Karmakar},\ and\ \citenamefont {Rakshit}}]{Pandey:2018wvh}%
  \BibitemOpen
  \bibfield  {author} {\bibinfo {author} {\bibfnamefont {S.}~\bibnamefont
  {Pandey}}, \bibinfo {author} {\bibfnamefont {S.}~\bibnamefont {Karmakar}}, \
  and\ \bibinfo {author} {\bibfnamefont {S.}~\bibnamefont {Rakshit}},\ }\href
  {\doibase 10.1007/JHEP11(2021)215} {\bibfield  {journal} {\bibinfo  {journal}
  {JHEP}\ }\textbf {\bibinfo {volume} {01}},\ \bibinfo {pages} {095} (\bibinfo
  {year} {2019})},\ \Eprint {http://arxiv.org/abs/1810.04203} {arXiv:1810.04203
  [hep-ph]} \BibitemShut {NoStop}%
\bibitem [{\citenamefont {Choi}\ \emph {et~al.}(2019)\citenamefont {Choi},
  \citenamefont {Kim},\ and\ \citenamefont {Rott}}]{Choi:2019ixb}%
  \BibitemOpen
  \bibfield  {author} {\bibinfo {author} {\bibfnamefont {K.-Y.}\ \bibnamefont
  {Choi}}, \bibinfo {author} {\bibfnamefont {J.}~\bibnamefont {Kim}}, \ and\
  \bibinfo {author} {\bibfnamefont {C.}~\bibnamefont {Rott}},\ }\href {\doibase
  10.1103/PhysRevD.99.083018} {\bibfield  {journal} {\bibinfo  {journal} {Phys.
  Rev. D}\ }\textbf {\bibinfo {volume} {99}},\ \bibinfo {pages} {083018}
  (\bibinfo {year} {2019})},\ \Eprint {http://arxiv.org/abs/1903.03302}
  {arXiv:1903.03302 [astro-ph.CO]} \BibitemShut {NoStop}%
\bibitem [{\citenamefont {Stadler}\ \emph {et~al.}(2019)\citenamefont
  {Stadler}, \citenamefont {B\oe{}hm},\ and\ \citenamefont
  {Mena}}]{Stadler:2019dii}%
  \BibitemOpen
  \bibfield  {author} {\bibinfo {author} {\bibfnamefont {J.}~\bibnamefont
  {Stadler}}, \bibinfo {author} {\bibfnamefont {C.}~\bibnamefont {B\oe{}hm}}, \
  and\ \bibinfo {author} {\bibfnamefont {O.}~\bibnamefont {Mena}},\ }\href
  {\doibase 10.1088/1475-7516/2019/08/014} {\bibfield  {journal} {\bibinfo
  {journal} {JCAP}\ }\textbf {\bibinfo {volume} {08}},\ \bibinfo {pages} {014}
  (\bibinfo {year} {2019})},\ \Eprint {http://arxiv.org/abs/1903.00540}
  {arXiv:1903.00540 [astro-ph.CO]} \BibitemShut {NoStop}%
\bibitem [{\citenamefont {Wilkinson}\ \emph {et~al.}(2014)\citenamefont
  {Wilkinson}, \citenamefont {Lesgourgues},\ and\ \citenamefont
  {Boehm}}]{Wilkinson:2013kia}%
  \BibitemOpen
  \bibfield  {author} {\bibinfo {author} {\bibfnamefont {R.~J.}\ \bibnamefont
  {Wilkinson}}, \bibinfo {author} {\bibfnamefont {J.}~\bibnamefont
  {Lesgourgues}}, \ and\ \bibinfo {author} {\bibfnamefont {C.}~\bibnamefont
  {Boehm}},\ }\href {\doibase 10.1088/1475-7516/2014/04/026} {\bibfield
  {journal} {\bibinfo  {journal} {JCAP}\ }\textbf {\bibinfo {volume} {04}},\
  \bibinfo {pages} {026} (\bibinfo {year} {2014})},\ \Eprint
  {http://arxiv.org/abs/1309.7588} {arXiv:1309.7588 [astro-ph.CO]} \BibitemShut
  {NoStop}%
\bibitem [{\citenamefont {Stadler}\ and\ \citenamefont
  {B\oe{}hm}(2018)}]{Stadler:2018jin}%
  \BibitemOpen
  \bibfield  {author} {\bibinfo {author} {\bibfnamefont {J.}~\bibnamefont
  {Stadler}}\ and\ \bibinfo {author} {\bibfnamefont {C.}~\bibnamefont
  {B\oe{}hm}},\ }\href {\doibase 10.1088/1475-7516/2018/10/009} {\bibfield
  {journal} {\bibinfo  {journal} {JCAP}\ }\textbf {\bibinfo {volume} {10}},\
  \bibinfo {pages} {009} (\bibinfo {year} {2018})},\ \Eprint
  {http://arxiv.org/abs/1802.06589} {arXiv:1802.06589 [astro-ph.CO]}
  \BibitemShut {NoStop}%
\bibitem [{\citenamefont {Becker}\ \emph {et~al.}(2021)\citenamefont {Becker},
  \citenamefont {Hooper}, \citenamefont {Kahlhoefer}, \citenamefont
  {Lesgourgues},\ and\ \citenamefont {Sch\"oneberg}}]{Becker:2020hzj}%
  \BibitemOpen
  \bibfield  {author} {\bibinfo {author} {\bibfnamefont {N.}~\bibnamefont
  {Becker}}, \bibinfo {author} {\bibfnamefont {D.~C.}\ \bibnamefont {Hooper}},
  \bibinfo {author} {\bibfnamefont {F.}~\bibnamefont {Kahlhoefer}}, \bibinfo
  {author} {\bibfnamefont {J.}~\bibnamefont {Lesgourgues}}, \ and\ \bibinfo
  {author} {\bibfnamefont {N.}~\bibnamefont {Sch\"oneberg}},\ }\href {\doibase
  10.1088/1475-7516/2021/02/019} {\bibfield  {journal} {\bibinfo  {journal}
  {JCAP}\ }\textbf {\bibinfo {volume} {02}},\ \bibinfo {pages} {019} (\bibinfo
  {year} {2021})},\ \Eprint {http://arxiv.org/abs/2010.04074} {arXiv:2010.04074
  [astro-ph.CO]} \BibitemShut {NoStop}%
\bibitem [{\citenamefont {Weiner}\ and\ \citenamefont
  {Yavin}(2012)}]{Weiner:2012cb}%
  \BibitemOpen
  \bibfield  {author} {\bibinfo {author} {\bibfnamefont {N.}~\bibnamefont
  {Weiner}}\ and\ \bibinfo {author} {\bibfnamefont {I.}~\bibnamefont {Yavin}},\
  }\href {\doibase 10.1103/PhysRevD.86.075021} {\bibfield  {journal} {\bibinfo
  {journal} {Phys. Rev. D}\ }\textbf {\bibinfo {volume} {86}},\ \bibinfo
  {pages} {075021} (\bibinfo {year} {2012})},\ \Eprint
  {http://arxiv.org/abs/1206.2910} {arXiv:1206.2910 [hep-ph]} \BibitemShut
  {NoStop}%
\bibitem [{\citenamefont {Boehm}\ \emph {et~al.}(2014)\citenamefont {Boehm},
  \citenamefont {Schewtschenko}, \citenamefont {Wilkinson}, \citenamefont
  {Baugh},\ and\ \citenamefont {Pascoli}}]{Boehm:2014vja}%
  \BibitemOpen
  \bibfield  {author} {\bibinfo {author} {\bibfnamefont {C.}~\bibnamefont
  {Boehm}}, \bibinfo {author} {\bibfnamefont {J.~A.}\ \bibnamefont
  {Schewtschenko}}, \bibinfo {author} {\bibfnamefont {R.~J.}\ \bibnamefont
  {Wilkinson}}, \bibinfo {author} {\bibfnamefont {C.~M.}\ \bibnamefont
  {Baugh}}, \ and\ \bibinfo {author} {\bibfnamefont {S.}~\bibnamefont
  {Pascoli}},\ }\href {\doibase 10.1093/mnrasl/slu115} {\bibfield  {journal}
  {\bibinfo  {journal} {Mon. Not. Roy. Astron. Soc.}\ }\textbf {\bibinfo
  {volume} {445}},\ \bibinfo {pages} {L31} (\bibinfo {year} {2014})},\ \Eprint
  {http://arxiv.org/abs/1404.7012} {arXiv:1404.7012 [astro-ph.CO]} \BibitemShut
  {NoStop}%
\bibitem [{\citenamefont {Kumar}\ \emph {et~al.}(2018)\citenamefont {Kumar},
  \citenamefont {Nunes},\ and\ \citenamefont {Yadav}}]{Kumar:2018yhh}%
  \BibitemOpen
  \bibfield  {author} {\bibinfo {author} {\bibfnamefont {S.}~\bibnamefont
  {Kumar}}, \bibinfo {author} {\bibfnamefont {R.~C.}\ \bibnamefont {Nunes}}, \
  and\ \bibinfo {author} {\bibfnamefont {S.~K.}\ \bibnamefont {Yadav}},\ }\href
  {\doibase 10.1103/PhysRevD.98.043521} {\bibfield  {journal} {\bibinfo
  {journal} {Phys. Rev. D}\ }\textbf {\bibinfo {volume} {98}},\ \bibinfo
  {pages} {043521} (\bibinfo {year} {2018})},\ \Eprint
  {http://arxiv.org/abs/1803.10229} {arXiv:1803.10229 [astro-ph.CO]}
  \BibitemShut {NoStop}%
\bibitem [{\citenamefont {Escudero}\ \emph {et~al.}(2018)\citenamefont
  {Escudero}, \citenamefont {Lopez-Honorez}, \citenamefont {Mena},
  \citenamefont {Palomares-Ruiz},\ and\ \citenamefont
  {Villanueva-Domingo}}]{Escudero:2018thh}%
  \BibitemOpen
  \bibfield  {author} {\bibinfo {author} {\bibfnamefont {M.}~\bibnamefont
  {Escudero}}, \bibinfo {author} {\bibfnamefont {L.}~\bibnamefont
  {Lopez-Honorez}}, \bibinfo {author} {\bibfnamefont {O.}~\bibnamefont {Mena}},
  \bibinfo {author} {\bibfnamefont {S.}~\bibnamefont {Palomares-Ruiz}}, \ and\
  \bibinfo {author} {\bibfnamefont {P.}~\bibnamefont {Villanueva-Domingo}},\
  }\href {\doibase 10.1088/1475-7516/2018/06/007} {\bibfield  {journal}
  {\bibinfo  {journal} {JCAP}\ }\textbf {\bibinfo {volume} {06}},\ \bibinfo
  {pages} {007} (\bibinfo {year} {2018})},\ \Eprint
  {http://arxiv.org/abs/1803.08427} {arXiv:1803.08427 [astro-ph.CO]}
  \BibitemShut {NoStop}%
\bibitem [{\citenamefont {Boehm}\ \emph {et~al.}(2013)\citenamefont {Boehm},
  \citenamefont {Dolan},\ and\ \citenamefont {McCabe}}]{Boehm:2013jpa}%
  \BibitemOpen
  \bibfield  {author} {\bibinfo {author} {\bibfnamefont {C.}~\bibnamefont
  {Boehm}}, \bibinfo {author} {\bibfnamefont {M.~J.}\ \bibnamefont {Dolan}}, \
  and\ \bibinfo {author} {\bibfnamefont {C.}~\bibnamefont {McCabe}},\ }\href
  {\doibase 10.1088/1475-7516/2013/08/041} {\bibfield  {journal} {\bibinfo
  {journal} {JCAP}\ }\textbf {\bibinfo {volume} {08}},\ \bibinfo {pages} {041}
  (\bibinfo {year} {2013})},\ \Eprint {http://arxiv.org/abs/1303.6270}
  {arXiv:1303.6270 [hep-ph]} \BibitemShut {NoStop}%
\bibitem [{\citenamefont {Boehm}\ \emph {et~al.}(2012)\citenamefont {Boehm},
  \citenamefont {Dolan},\ and\ \citenamefont {McCabe}}]{Boehm:2012gr}%
  \BibitemOpen
  \bibfield  {author} {\bibinfo {author} {\bibfnamefont {C.}~\bibnamefont
  {Boehm}}, \bibinfo {author} {\bibfnamefont {M.~J.}\ \bibnamefont {Dolan}}, \
  and\ \bibinfo {author} {\bibfnamefont {C.}~\bibnamefont {McCabe}},\ }\href
  {\doibase 10.1088/1475-7516/2012/12/027} {\bibfield  {journal} {\bibinfo
  {journal} {JCAP}\ }\textbf {\bibinfo {volume} {12}},\ \bibinfo {pages} {027}
  (\bibinfo {year} {2012})},\ \Eprint {http://arxiv.org/abs/1207.0497}
  {arXiv:1207.0497 [astro-ph.CO]} \BibitemShut {NoStop}%
\bibitem [{\citenamefont {Ho}\ and\ \citenamefont
  {Scherrer}(2013)}]{Ho:2012ug}%
  \BibitemOpen
  \bibfield  {author} {\bibinfo {author} {\bibfnamefont {C.~M.}\ \bibnamefont
  {Ho}}\ and\ \bibinfo {author} {\bibfnamefont {R.~J.}\ \bibnamefont
  {Scherrer}},\ }\href {\doibase 10.1103/PhysRevD.87.023505} {\bibfield
  {journal} {\bibinfo  {journal} {Phys. Rev. D}\ }\textbf {\bibinfo {volume}
  {87}},\ \bibinfo {pages} {023505} (\bibinfo {year} {2013})},\ \Eprint
  {http://arxiv.org/abs/1208.4347} {arXiv:1208.4347 [astro-ph.CO]} \BibitemShut
  {NoStop}%
\bibitem [{\citenamefont {Steigman}(2013)}]{Steigman:2013yua}%
  \BibitemOpen
  \bibfield  {author} {\bibinfo {author} {\bibfnamefont {G.}~\bibnamefont
  {Steigman}},\ }\href {\doibase 10.1103/PhysRevD.87.103517} {\bibfield
  {journal} {\bibinfo  {journal} {Phys. Rev. D}\ }\textbf {\bibinfo {volume}
  {87}},\ \bibinfo {pages} {103517} (\bibinfo {year} {2013})},\ \Eprint
  {http://arxiv.org/abs/1303.0049} {arXiv:1303.0049 [astro-ph.CO]} \BibitemShut
  {NoStop}%
\bibitem [{\citenamefont {Nollett}\ and\ \citenamefont
  {Steigman}(2015)}]{Nollett:2014lwa}%
  \BibitemOpen
  \bibfield  {author} {\bibinfo {author} {\bibfnamefont {K.~M.}\ \bibnamefont
  {Nollett}}\ and\ \bibinfo {author} {\bibfnamefont {G.}~\bibnamefont
  {Steigman}},\ }\href {\doibase 10.1103/PhysRevD.91.083505} {\bibfield
  {journal} {\bibinfo  {journal} {Phys. Rev. D}\ }\textbf {\bibinfo {volume}
  {91}},\ \bibinfo {pages} {083505} (\bibinfo {year} {2015})},\ \Eprint
  {http://arxiv.org/abs/1411.6005} {arXiv:1411.6005 [astro-ph.CO]} \BibitemShut
  {NoStop}%
\bibitem [{\citenamefont {Steigman}\ and\ \citenamefont
  {Nollett}(2015)}]{Steigman:2014uqa}%
  \BibitemOpen
  \bibfield  {author} {\bibinfo {author} {\bibfnamefont {G.}~\bibnamefont
  {Steigman}}\ and\ \bibinfo {author} {\bibfnamefont {K.~M.}\ \bibnamefont
  {Nollett}},\ }\href {\doibase 10.1016/j.phpro.2014.12.029} {\bibfield
  {journal} {\bibinfo  {journal} {Phys. Procedia}\ }\textbf {\bibinfo {volume}
  {61}},\ \bibinfo {pages} {179} (\bibinfo {year} {2015})},\ \Eprint
  {http://arxiv.org/abs/1402.5399} {arXiv:1402.5399 [astro-ph.CO]} \BibitemShut
  {NoStop}%
\bibitem [{\citenamefont {Serpico}\ and\ \citenamefont
  {Raffelt}(2004)}]{Serpico:2004nm}%
  \BibitemOpen
  \bibfield  {author} {\bibinfo {author} {\bibfnamefont {P.~D.}\ \bibnamefont
  {Serpico}}\ and\ \bibinfo {author} {\bibfnamefont {G.~G.}\ \bibnamefont
  {Raffelt}},\ }\href {\doibase 10.1103/PhysRevD.70.043526} {\bibfield
  {journal} {\bibinfo  {journal} {Phys. Rev. D}\ }\textbf {\bibinfo {volume}
  {70}},\ \bibinfo {pages} {043526} (\bibinfo {year} {2004})},\ \Eprint
  {http://arxiv.org/abs/astro-ph/0403417} {arXiv:astro-ph/0403417} \BibitemShut
  {NoStop}%
\bibitem [{\citenamefont {Hochberg}\ \emph
  {et~al.}(2016{\natexlab{a}})\citenamefont {Hochberg}, \citenamefont {Pyle},
  \citenamefont {Zhao},\ and\ \citenamefont {Zurek}}]{Hochberg:2015fth}%
  \BibitemOpen
  \bibfield  {author} {\bibinfo {author} {\bibfnamefont {Y.}~\bibnamefont
  {Hochberg}}, \bibinfo {author} {\bibfnamefont {M.}~\bibnamefont {Pyle}},
  \bibinfo {author} {\bibfnamefont {Y.}~\bibnamefont {Zhao}}, \ and\ \bibinfo
  {author} {\bibfnamefont {K.~M.}\ \bibnamefont {Zurek}},\ }\href {\doibase
  10.1007/JHEP08(2016)057} {\bibfield  {journal} {\bibinfo  {journal} {JHEP}\
  }\textbf {\bibinfo {volume} {08}},\ \bibinfo {pages} {057} (\bibinfo {year}
  {2016}{\natexlab{a}})},\ \Eprint {http://arxiv.org/abs/1512.04533}
  {arXiv:1512.04533 [hep-ph]} \BibitemShut {NoStop}%
\bibitem [{\citenamefont {Hochberg}\ \emph
  {et~al.}(2016{\natexlab{b}})\citenamefont {Hochberg}, \citenamefont {Zhao},\
  and\ \citenamefont {Zurek}}]{Hochberg:2015pha}%
  \BibitemOpen
  \bibfield  {author} {\bibinfo {author} {\bibfnamefont {Y.}~\bibnamefont
  {Hochberg}}, \bibinfo {author} {\bibfnamefont {Y.}~\bibnamefont {Zhao}}, \
  and\ \bibinfo {author} {\bibfnamefont {K.~M.}\ \bibnamefont {Zurek}},\ }\href
  {\doibase 10.1103/PhysRevLett.116.011301} {\bibfield  {journal} {\bibinfo
  {journal} {Phys. Rev. Lett.}\ }\textbf {\bibinfo {volume} {116}},\ \bibinfo
  {pages} {011301} (\bibinfo {year} {2016}{\natexlab{b}})},\ \Eprint
  {http://arxiv.org/abs/1504.07237} {arXiv:1504.07237 [hep-ph]} \BibitemShut
  {NoStop}%
\bibitem [{\citenamefont {Schutz}\ and\ \citenamefont
  {Zurek}(2016)}]{Schutz:2016tid}%
  \BibitemOpen
  \bibfield  {author} {\bibinfo {author} {\bibfnamefont {K.}~\bibnamefont
  {Schutz}}\ and\ \bibinfo {author} {\bibfnamefont {K.~M.}\ \bibnamefont
  {Zurek}},\ }\href {\doibase 10.1103/PhysRevLett.117.121302} {\bibfield
  {journal} {\bibinfo  {journal} {Phys. Rev. Lett.}\ }\textbf {\bibinfo
  {volume} {117}},\ \bibinfo {pages} {121302} (\bibinfo {year} {2016})},\
  \Eprint {http://arxiv.org/abs/1604.08206} {arXiv:1604.08206 [hep-ph]}
  \BibitemShut {NoStop}%
\bibitem [{\citenamefont {Knapen}\ \emph {et~al.}(2017)\citenamefont {Knapen},
  \citenamefont {Lin},\ and\ \citenamefont {Zurek}}]{Knapen:2016cue}%
  \BibitemOpen
  \bibfield  {author} {\bibinfo {author} {\bibfnamefont {S.}~\bibnamefont
  {Knapen}}, \bibinfo {author} {\bibfnamefont {T.}~\bibnamefont {Lin}}, \ and\
  \bibinfo {author} {\bibfnamefont {K.~M.}\ \bibnamefont {Zurek}},\ }\href
  {\doibase 10.1103/PhysRevD.95.056019} {\bibfield  {journal} {\bibinfo
  {journal} {Phys. Rev. D}\ }\textbf {\bibinfo {volume} {95}},\ \bibinfo
  {pages} {056019} (\bibinfo {year} {2017})},\ \Eprint
  {http://arxiv.org/abs/1611.06228} {arXiv:1611.06228 [hep-ph]} \BibitemShut
  {NoStop}%
\bibitem [{\citenamefont {Hu}(1998)}]{Hu:1998kj}%
  \BibitemOpen
  \bibfield  {author} {\bibinfo {author} {\bibfnamefont {W.}~\bibnamefont
  {Hu}},\ }\href {\doibase 10.1086/306274} {\bibfield  {journal} {\bibinfo
  {journal} {Astrophys. J.}\ }\textbf {\bibinfo {volume} {506}},\ \bibinfo
  {pages} {485} (\bibinfo {year} {1998})},\ \Eprint
  {http://arxiv.org/abs/astro-ph/9801234} {arXiv:astro-ph/9801234} \BibitemShut
  {NoStop}%
\bibitem [{\citenamefont {Thomas}\ \emph {et~al.}(2016)\citenamefont {Thomas},
  \citenamefont {Kopp},\ and\ \citenamefont {Skordis}}]{Thomas:2016iav}%
  \BibitemOpen
  \bibfield  {author} {\bibinfo {author} {\bibfnamefont {D.~B.}\ \bibnamefont
  {Thomas}}, \bibinfo {author} {\bibfnamefont {M.}~\bibnamefont {Kopp}}, \ and\
  \bibinfo {author} {\bibfnamefont {C.}~\bibnamefont {Skordis}},\ }\href
  {\doibase 10.3847/0004-637X/830/2/155} {\bibfield  {journal} {\bibinfo
  {journal} {Astrophys. J.}\ }\textbf {\bibinfo {volume} {830}},\ \bibinfo
  {pages} {155} (\bibinfo {year} {2016})},\ \Eprint
  {http://arxiv.org/abs/1601.05097} {arXiv:1601.05097 [astro-ph.CO]}
  \BibitemShut {NoStop}%
\bibitem [{\citenamefont {Kopp}\ \emph {et~al.}(2018)\citenamefont {Kopp},
  \citenamefont {Skordis}, \citenamefont {Thomas},\ and\ \citenamefont
  {Ili\'c}}]{Kopp:2018zxp}%
  \BibitemOpen
  \bibfield  {author} {\bibinfo {author} {\bibfnamefont {M.}~\bibnamefont
  {Kopp}}, \bibinfo {author} {\bibfnamefont {C.}~\bibnamefont {Skordis}},
  \bibinfo {author} {\bibfnamefont {D.~B.}\ \bibnamefont {Thomas}}, \ and\
  \bibinfo {author} {\bibfnamefont {S.}~\bibnamefont {Ili\'c}},\ }\href
  {\doibase 10.1103/PhysRevLett.120.221102} {\bibfield  {journal} {\bibinfo
  {journal} {Phys. Rev. Lett.}\ }\textbf {\bibinfo {volume} {120}},\ \bibinfo
  {pages} {221102} (\bibinfo {year} {2018})},\ \Eprint
  {http://arxiv.org/abs/1802.09541} {arXiv:1802.09541 [astro-ph.CO]}
  \BibitemShut {NoStop}%
\bibitem [{\citenamefont {Velten}\ \emph {et~al.}(2021)\citenamefont {Velten},
  \citenamefont {Costa},\ and\ \citenamefont {Zimdahl}}]{Velten:2021cqj}%
  \BibitemOpen
  \bibfield  {author} {\bibinfo {author} {\bibfnamefont {H.}~\bibnamefont
  {Velten}}, \bibinfo {author} {\bibfnamefont {I.}~\bibnamefont {Costa}}, \
  and\ \bibinfo {author} {\bibfnamefont {W.}~\bibnamefont {Zimdahl}},\
  }\href@noop {} {\  (\bibinfo {year} {2021})},\ \Eprint
  {http://arxiv.org/abs/2104.05352} {arXiv:2104.05352 [astro-ph.CO]}
  \BibitemShut {NoStop}%
\bibitem [{\citenamefont {Zimdahl}(1996)}]{Zimdahl:1996fj}%
  \BibitemOpen
  \bibfield  {author} {\bibinfo {author} {\bibfnamefont {W.}~\bibnamefont
  {Zimdahl}},\ }\href {\doibase 10.1093/mnras/280.4.1239} {\bibfield  {journal}
  {\bibinfo  {journal} {Mon. Not. Roy. Astron. Soc.}\ }\textbf {\bibinfo
  {volume} {280}},\ \bibinfo {pages} {1239} (\bibinfo {year} {1996})},\ \Eprint
  {http://arxiv.org/abs/astro-ph/9602128} {arXiv:astro-ph/9602128} \BibitemShut
  {NoStop}%
\bibitem [{\citenamefont {Viel}\ \emph {et~al.}(2013)\citenamefont {Viel},
  \citenamefont {Becker}, \citenamefont {Bolton},\ and\ \citenamefont
  {Haehnelt}}]{Viel:2013fqw}%
  \BibitemOpen
  \bibfield  {author} {\bibinfo {author} {\bibfnamefont {M.}~\bibnamefont
  {Viel}}, \bibinfo {author} {\bibfnamefont {G.~D.}\ \bibnamefont {Becker}},
  \bibinfo {author} {\bibfnamefont {J.~S.}\ \bibnamefont {Bolton}}, \ and\
  \bibinfo {author} {\bibfnamefont {M.~G.}\ \bibnamefont {Haehnelt}},\ }\href
  {\doibase 10.1103/PhysRevD.88.043502} {\bibfield  {journal} {\bibinfo
  {journal} {Phys. Rev. D}\ }\textbf {\bibinfo {volume} {88}},\ \bibinfo
  {pages} {043502} (\bibinfo {year} {2013})},\ \Eprint
  {http://arxiv.org/abs/1306.2314} {arXiv:1306.2314 [astro-ph.CO]} \BibitemShut
  {NoStop}%
\bibitem [{\citenamefont {Ma}\ and\ \citenamefont
  {Bertschinger}(1995)}]{Ma:1995ey}%
  \BibitemOpen
  \bibfield  {author} {\bibinfo {author} {\bibfnamefont {C.-P.}\ \bibnamefont
  {Ma}}\ and\ \bibinfo {author} {\bibfnamefont {E.}~\bibnamefont
  {Bertschinger}},\ }\href {\doibase 10.1086/176550} {\bibfield  {journal}
  {\bibinfo  {journal} {Astrophys. J.}\ }\textbf {\bibinfo {volume} {455}},\
  \bibinfo {pages} {7} (\bibinfo {year} {1995})},\ \Eprint
  {http://arxiv.org/abs/astro-ph/9506072} {arXiv:astro-ph/9506072} \BibitemShut
  {NoStop}%
\bibitem [{\citenamefont {Kazantzidis}\ and\ \citenamefont
  {Perivolaropoulos}(2018)}]{Kazantzidis:2018rnb}%
  \BibitemOpen
  \bibfield  {author} {\bibinfo {author} {\bibfnamefont {L.}~\bibnamefont
  {Kazantzidis}}\ and\ \bibinfo {author} {\bibfnamefont {L.}~\bibnamefont
  {Perivolaropoulos}},\ }\href {\doibase 10.1103/PhysRevD.97.103503} {\bibfield
   {journal} {\bibinfo  {journal} {Phys. Rev. D}\ }\textbf {\bibinfo {volume}
  {97}},\ \bibinfo {pages} {103503} (\bibinfo {year} {2018})},\ \Eprint
  {http://arxiv.org/abs/1803.01337} {arXiv:1803.01337 [astro-ph.CO]}
  \BibitemShut {NoStop}%
\bibitem [{\citenamefont {Alcock}\ and\ \citenamefont
  {Paczynski}(1979)}]{Alcock:1979mp}%
  \BibitemOpen
  \bibfield  {author} {\bibinfo {author} {\bibfnamefont {C.}~\bibnamefont
  {Alcock}}\ and\ \bibinfo {author} {\bibfnamefont {B.}~\bibnamefont
  {Paczynski}},\ }\href {\doibase 10.1038/281358a0} {\bibfield  {journal}
  {\bibinfo  {journal} {Nature}\ }\textbf {\bibinfo {volume} {281}},\ \bibinfo
  {pages} {358} (\bibinfo {year} {1979})}\BibitemShut {NoStop}%
\bibitem [{\citenamefont {Macaulay}\ \emph {et~al.}(2013)\citenamefont
  {Macaulay}, \citenamefont {Wehus},\ and\ \citenamefont
  {Eriksen}}]{Macaulay:2013swa}%
  \BibitemOpen
  \bibfield  {author} {\bibinfo {author} {\bibfnamefont {E.}~\bibnamefont
  {Macaulay}}, \bibinfo {author} {\bibfnamefont {I.~K.}\ \bibnamefont {Wehus}},
  \ and\ \bibinfo {author} {\bibfnamefont {H.~K.}\ \bibnamefont {Eriksen}},\
  }\href {\doibase 10.1103/PhysRevLett.111.161301} {\bibfield  {journal}
  {\bibinfo  {journal} {Phys. Rev. Lett.}\ }\textbf {\bibinfo {volume} {111}},\
  \bibinfo {pages} {161301} (\bibinfo {year} {2013})},\ \Eprint
  {http://arxiv.org/abs/1303.6583} {arXiv:1303.6583 [astro-ph.CO]} \BibitemShut
  {NoStop}%
\bibitem [{\citenamefont {Dodelson}\ \emph {et~al.}(1996)\citenamefont
  {Dodelson}, \citenamefont {Gates},\ and\ \citenamefont
  {Stebbins}}]{Dodelson:1995es}%
  \BibitemOpen
  \bibfield  {author} {\bibinfo {author} {\bibfnamefont {S.}~\bibnamefont
  {Dodelson}}, \bibinfo {author} {\bibfnamefont {E.}~\bibnamefont {Gates}}, \
  and\ \bibinfo {author} {\bibfnamefont {A.}~\bibnamefont {Stebbins}},\ }\href
  {\doibase 10.1086/177581} {\bibfield  {journal} {\bibinfo  {journal}
  {Astrophys. J.}\ }\textbf {\bibinfo {volume} {467}},\ \bibinfo {pages} {10}
  (\bibinfo {year} {1996})},\ \Eprint {http://arxiv.org/abs/astro-ph/9509147}
  {arXiv:astro-ph/9509147} \BibitemShut {NoStop}%
\bibitem [{\citenamefont {Sachs}\ and\ \citenamefont
  {Wolfe}(1967)}]{Sachs:1967er}%
  \BibitemOpen
  \bibfield  {author} {\bibinfo {author} {\bibfnamefont {R.~K.}\ \bibnamefont
  {Sachs}}\ and\ \bibinfo {author} {\bibfnamefont {A.~M.}\ \bibnamefont
  {Wolfe}},\ }\href {\doibase 10.1007/s10714-007-0448-9} {\bibfield  {journal}
  {\bibinfo  {journal} {Astrophys. J.}\ }\textbf {\bibinfo {volume} {147}},\
  \bibinfo {pages} {73} (\bibinfo {year} {1967})}\BibitemShut {NoStop}%
\bibitem [{\citenamefont {Vagnozzi}(2021)}]{Vagnozzi:2021gjh}%
  \BibitemOpen
  \bibfield  {author} {\bibinfo {author} {\bibfnamefont {S.}~\bibnamefont
  {Vagnozzi}},\ }\href {\doibase 10.1103/PhysRevD.104.063524} {\bibfield
  {journal} {\bibinfo  {journal} {Phys. Rev. D}\ }\textbf {\bibinfo {volume}
  {104}},\ \bibinfo {pages} {063524} (\bibinfo {year} {2021})},\ \Eprint
  {http://arxiv.org/abs/2105.10425} {arXiv:2105.10425 [astro-ph.CO]}
  \BibitemShut {NoStop}%
\bibitem [{\citenamefont {Lesgourgues}\ and\ \citenamefont
  {Pastor}(2014)}]{Lesgourgues:2014zoa}%
  \BibitemOpen
  \bibfield  {author} {\bibinfo {author} {\bibfnamefont {J.}~\bibnamefont
  {Lesgourgues}}\ and\ \bibinfo {author} {\bibfnamefont {S.}~\bibnamefont
  {Pastor}},\ }\href {\doibase 10.1088/1367-2630/16/6/065002} {\bibfield
  {journal} {\bibinfo  {journal} {New J. Phys.}\ }\textbf {\bibinfo {volume}
  {16}},\ \bibinfo {pages} {065002} (\bibinfo {year} {2014})},\ \Eprint
  {http://arxiv.org/abs/1404.1740} {arXiv:1404.1740 [hep-ph]} \BibitemShut
  {NoStop}%
\bibitem [{\citenamefont {Lesgourgues}(2011)}]{Lesgourgues:2011re}%
  \BibitemOpen
  \bibfield  {author} {\bibinfo {author} {\bibfnamefont {J.}~\bibnamefont
  {Lesgourgues}},\ }\href@noop {} {\  (\bibinfo {year} {2011})},\ \Eprint
  {http://arxiv.org/abs/1104.2932} {arXiv:1104.2932 [astro-ph.IM]} \BibitemShut
  {NoStop}%
\bibitem [{\citenamefont {Blas}\ \emph {et~al.}(2011)\citenamefont {Blas},
  \citenamefont {Lesgourgues},\ and\ \citenamefont {Tram}}]{Blas:2011rf}%
  \BibitemOpen
  \bibfield  {author} {\bibinfo {author} {\bibfnamefont {D.}~\bibnamefont
  {Blas}}, \bibinfo {author} {\bibfnamefont {J.}~\bibnamefont {Lesgourgues}}, \
  and\ \bibinfo {author} {\bibfnamefont {T.}~\bibnamefont {Tram}},\ }\href
  {\doibase 10.1088/1475-7516/2011/07/034} {\bibfield  {journal} {\bibinfo
  {journal} {JCAP}\ }\textbf {\bibinfo {volume} {07}},\ \bibinfo {pages} {034}
  (\bibinfo {year} {2011})},\ \Eprint {http://arxiv.org/abs/1104.2933}
  {arXiv:1104.2933 [astro-ph.CO]} \BibitemShut {NoStop}%
\bibitem [{\citenamefont {Mead}\ \emph {et~al.}(2015)\citenamefont {Mead},
  \citenamefont {Peacock}, \citenamefont {Heymans}, \citenamefont {Joudaki},\
  and\ \citenamefont {Heavens}}]{Mead:2015yca}%
  \BibitemOpen
  \bibfield  {author} {\bibinfo {author} {\bibfnamefont {A.}~\bibnamefont
  {Mead}}, \bibinfo {author} {\bibfnamefont {J.}~\bibnamefont {Peacock}},
  \bibinfo {author} {\bibfnamefont {C.}~\bibnamefont {Heymans}}, \bibinfo
  {author} {\bibfnamefont {S.}~\bibnamefont {Joudaki}}, \ and\ \bibinfo
  {author} {\bibfnamefont {A.}~\bibnamefont {Heavens}},\ }\href {\doibase
  10.1093/mnras/stv2036} {\bibfield  {journal} {\bibinfo  {journal} {Mon. Not.
  Roy. Astron. Soc.}\ }\textbf {\bibinfo {volume} {454}},\ \bibinfo {pages}
  {1958} (\bibinfo {year} {2015})},\ \Eprint {http://arxiv.org/abs/1505.07833}
  {arXiv:1505.07833 [astro-ph.CO]} \BibitemShut {NoStop}%
\bibitem [{\citenamefont {Mead}(2017)}]{Mead:2016ybv}%
  \BibitemOpen
  \bibfield  {author} {\bibinfo {author} {\bibfnamefont {A.}~\bibnamefont
  {Mead}},\ }\href {\doibase 10.1093/mnras/stw2312} {\bibfield  {journal}
  {\bibinfo  {journal} {Mon. Not. Roy. Astron. Soc.}\ }\textbf {\bibinfo
  {volume} {464}},\ \bibinfo {pages} {1282} (\bibinfo {year} {2017})},\ \Eprint
  {http://arxiv.org/abs/1606.05345} {arXiv:1606.05345 [astro-ph.CO]}
  \BibitemShut {NoStop}%
\bibitem [{\citenamefont {Mead}\ \emph {et~al.}(2020)\citenamefont {Mead},
  \citenamefont {Brieden}, \citenamefont {Tr\"oster},\ and\ \citenamefont
  {Heymans}}]{Mead:2020vgs}%
  \BibitemOpen
  \bibfield  {author} {\bibinfo {author} {\bibfnamefont {A.}~\bibnamefont
  {Mead}}, \bibinfo {author} {\bibfnamefont {S.}~\bibnamefont {Brieden}},
  \bibinfo {author} {\bibfnamefont {T.}~\bibnamefont {Tr\"oster}}, \ and\
  \bibinfo {author} {\bibfnamefont {C.}~\bibnamefont {Heymans}},\ }\href
  {\doibase 10.1093/mnras/stab082} {\  (\bibinfo {year} {2020}),\
  10.1093/mnras/stab082},\ \Eprint {http://arxiv.org/abs/2009.01858}
  {arXiv:2009.01858 [astro-ph.CO]} \BibitemShut {NoStop}%
\bibitem [{\citenamefont {Torrado}\ and\ \citenamefont
  {Lewis}(2020)}]{Torrado:2020dgo}%
  \BibitemOpen
  \bibfield  {author} {\bibinfo {author} {\bibfnamefont {J.}~\bibnamefont
  {Torrado}}\ and\ \bibinfo {author} {\bibfnamefont {A.}~\bibnamefont
  {Lewis}},\ }\href@noop {} {\  (\bibinfo {year} {2020})},\ \Eprint
  {http://arxiv.org/abs/2005.05290} {arXiv:2005.05290 [astro-ph.IM]}
  \BibitemShut {NoStop}%
\bibitem [{\citenamefont {Gelman}\ and\ \citenamefont
  {Rubin}(1992)}]{10.1214/ss/1177011136}%
  \BibitemOpen
  \bibfield  {author} {\bibinfo {author} {\bibfnamefont {A.}~\bibnamefont
  {Gelman}}\ and\ \bibinfo {author} {\bibfnamefont {D.~B.}\ \bibnamefont
  {Rubin}},\ }\href {\doibase 10.1214/ss/1177011136} {\bibfield  {journal}
  {\bibinfo  {journal} {Statistical Science}\ }\textbf {\bibinfo {volume}
  {7}},\ \bibinfo {pages} {457 } (\bibinfo {year} {1992})}\BibitemShut
  {NoStop}%
\bibitem [{\citenamefont {Lewis}(2019)}]{Lewis:2019xzd}%
  \BibitemOpen
  \bibfield  {author} {\bibinfo {author} {\bibfnamefont {A.}~\bibnamefont
  {Lewis}},\ }\href@noop {} {\  (\bibinfo {year} {2019})},\ \Eprint
  {http://arxiv.org/abs/1910.13970} {arXiv:1910.13970 [astro-ph.IM]}
  \BibitemShut {NoStop}%
\bibitem [{\citenamefont {Aghanim}\ \emph
  {et~al.}(2020{\natexlab{b}})\citenamefont {Aghanim} \emph
  {et~al.}}]{Planck:2018lbu}%
  \BibitemOpen
  \bibfield  {author} {\bibinfo {author} {\bibfnamefont {N.}~\bibnamefont
  {Aghanim}} \emph {et~al.} (\bibinfo {collaboration} {Planck}),\ }\href
  {\doibase 10.1051/0004-6361/201833886} {\bibfield  {journal} {\bibinfo
  {journal} {Astron. Astrophys.}\ }\textbf {\bibinfo {volume} {641}},\ \bibinfo
  {pages} {A8} (\bibinfo {year} {2020}{\natexlab{b}})},\ \Eprint
  {http://arxiv.org/abs/1807.06210} {arXiv:1807.06210 [astro-ph.CO]}
  \BibitemShut {NoStop}%
\bibitem [{\citenamefont {Riess}\ \emph {et~al.}(2021)\citenamefont {Riess},
  \citenamefont {Casertano}, \citenamefont {Yuan}, \citenamefont {Bowers},
  \citenamefont {Macri}, \citenamefont {Zinn},\ and\ \citenamefont
  {Scolnic}}]{Riess:2020fzl}%
  \BibitemOpen
  \bibfield  {author} {\bibinfo {author} {\bibfnamefont {A.~G.}\ \bibnamefont
  {Riess}}, \bibinfo {author} {\bibfnamefont {S.}~\bibnamefont {Casertano}},
  \bibinfo {author} {\bibfnamefont {W.}~\bibnamefont {Yuan}}, \bibinfo {author}
  {\bibfnamefont {J.~B.}\ \bibnamefont {Bowers}}, \bibinfo {author}
  {\bibfnamefont {L.}~\bibnamefont {Macri}}, \bibinfo {author} {\bibfnamefont
  {J.~C.}\ \bibnamefont {Zinn}}, \ and\ \bibinfo {author} {\bibfnamefont
  {D.}~\bibnamefont {Scolnic}},\ }\href {\doibase 10.3847/2041-8213/abdbaf}
  {\bibfield  {journal} {\bibinfo  {journal} {Astrophys. J. Lett.}\ }\textbf
  {\bibinfo {volume} {908}},\ \bibinfo {pages} {L6} (\bibinfo {year} {2021})},\
  \Eprint {http://arxiv.org/abs/2012.08534} {arXiv:2012.08534 [astro-ph.CO]}
  \BibitemShut {NoStop}%
\bibitem [{\citenamefont {Hill}\ \emph {et~al.}(2020)\citenamefont {Hill},
  \citenamefont {McDonough}, \citenamefont {Toomey},\ and\ \citenamefont
  {Alexander}}]{Hill:2020osr}%
  \BibitemOpen
  \bibfield  {author} {\bibinfo {author} {\bibfnamefont {J.~C.}\ \bibnamefont
  {Hill}}, \bibinfo {author} {\bibfnamefont {E.}~\bibnamefont {McDonough}},
  \bibinfo {author} {\bibfnamefont {M.~W.}\ \bibnamefont {Toomey}}, \ and\
  \bibinfo {author} {\bibfnamefont {S.}~\bibnamefont {Alexander}},\ }\href
  {\doibase 10.1103/PhysRevD.102.043507} {\bibfield  {journal} {\bibinfo
  {journal} {Phys. Rev. D}\ }\textbf {\bibinfo {volume} {102}},\ \bibinfo
  {pages} {043507} (\bibinfo {year} {2020})},\ \Eprint
  {http://arxiv.org/abs/2003.07355} {arXiv:2003.07355 [astro-ph.CO]}
  \BibitemShut {NoStop}%
\bibitem [{\citenamefont {Aghanim}\ \emph {et~al.}(2016)\citenamefont {Aghanim}
  \emph {et~al.}}]{Planck:2015bpv}%
  \BibitemOpen
  \bibfield  {author} {\bibinfo {author} {\bibfnamefont {N.}~\bibnamefont
  {Aghanim}} \emph {et~al.} (\bibinfo {collaboration} {Planck}),\ }\href
  {\doibase 10.1051/0004-6361/201526926} {\bibfield  {journal} {\bibinfo
  {journal} {Astron. Astrophys.}\ }\textbf {\bibinfo {volume} {594}},\ \bibinfo
  {pages} {A11} (\bibinfo {year} {2016})},\ \Eprint
  {http://arxiv.org/abs/1507.02704} {arXiv:1507.02704 [astro-ph.CO]}
  \BibitemShut {NoStop}%
\end{thebibliography}%

\end{document}